\documentclass[twoside,aps,prl,twocolumn,superscriptaddress,showpacs]{revtex4-1}

\usepackage{amsmath}
\usepackage{amssymb}
\usepackage{graphicx}
\usepackage{esint}

\usepackage{amsfonts}
\usepackage{subfiles}

\begin{document}

\title{Voltage noise, switching rates, and multiple phase-slips in moderately
damped Josephson junctions}

\author{Martin \v{Z}onda}

\affiliation{Department of Condensed Matter Physics, Faculty of Mathematics and
Physics, Charles University in Prague, Ke Karlovu 5, 121 16 Praha
2, Czech Republic}

\author{Wolfgang Belzig}

\affiliation{Fachbereich Physik, Universit\"{a}t Konstanz, D-78457 Konstanz, Germany}

\author{Tom\'{a}\v{s} Novotn\'{y} }

\email{tno@karlov.mff.cuni.cz}
\affiliation{Department of Condensed Matter Physics, Faculty of Mathematics and
Physics, Charles University in Prague, Ke Karlovu 5, 121 16 Praha
2, Czech Republic}

\date{\today}
\begin{abstract}
We study the voltage noise properties including the switching rates
and statistics of phase-slips in moderately damped Josephson junctions
using a novel efficient numerical approach combining the matrix continued-fraction
method with the full counting statistics. By analyzing the noise results
obtained for the RCSJ model we identify different dominating components,
namely the thermal noise close to equilibrium (small current-bias
regime), the shot noise of (multiple) phase-slips in the intermediate
range of biases and the switching noise for yet higher bias currents.
We extract thus far inaccessible characteristic rates of phase-slips
in the shot noise regime as well as the escape and retrapping rates
in the switching regime as functions of various junction's parameters.
The method can be extended and applied to other experimentally relevant
Josephson junction circuits.
\end{abstract}


\maketitle
\emph{Introduction.--- }Josephson junctions (JJs) and their dynamics
have been subject of intensive study ever since the discovery of the
Josephson effect \cite{Barone,  *Likharev86} not only because of their
fascinating microscopic physical properties and high application potential
but also for their ability to implement via phase dynamics elementary
concepts of nonlinear dynamical systems such as chaos, multistability,
and switching \cite{Kautz96,Risken}. In recent years there has been
progress in fabricating unconventional meso- \cite{Budzin05,Sickinger12}
and nanoscopic \cite{De-Franceschi10} JJs exhibiting among other
effects nonsinusoidal current-phase relations with exotic dynamics
\cite{Goldobin:PRL13}. However, small junctions are more prone to
the influence of environmental noise and their dynamics is inherently
stochastic \cite{Kautz89,Kautz90}. Due to the richness of dynamical
regimes even the description of conventional junctions especially
in the intermediate damping regime may be challenging and one has
to often resort to lengthy simulations \cite{Fenton08}. 

Another side-effect of miniaturization connected to the stochastic
nature of the problem is the shift of interest from just mean quantities
such as the mean voltage to more elaborate statistical description
including, e.g., the voltage noise in simulations \cite{Voss81},
theory \cite{Golubev10,Zonda:PhScr12}, as well as experiment %
\footnote{P. Hakonen, private communication.%
}. In the present paper we introduce a robust and efficient numerical
method based on the matrix continued-fraction (MCF) method \cite{Risken}
which can be used to study the voltage noise of JJs with an arbitrary
level of the phase dynamics damping. The method reveals various regimes
of current-biasing the junction with the corresponding dominant voltage-noise
mechanisms, including the thermal noise, multiple phase-slips (MPS),
and switching processes with related escape and retrapping rates whose
values are easily determined. When combined with the full counting
statistics (FCS) of the phase dynamics \cite{Golubev10} it allows
us to decompose (together with a clear verification mechanism, when
the decomposition is legitimate) the phase dynamics into independent
elementary processes \cite{Vanevic:PRL07, *Vanevic:PRB08, *Padurariu:PRB12}
constituted by MPS and find their rates. 

We demonstrate the method on the paradigmatic case of the RCSJ model.
However, after appropriate minor extension it is also applicable to
other JJ models providing further novel results such as the voltage
noise in JJs with arbitrary current-phase relations as in Ref.~\cite{Sickinger12},
description of the experimentally-relevant circuits with structured
electromagnetic environments \cite{Joyez99,Zonda:PhScr12}, and/or
frequency-dependent voltage noise for these models. Moreover, due
to the phase-charge duality our results are also directly relevant
for the current-noise in the nanowire quantum phase-slip circuits
\cite{Webster13}. We defer discussion of these new results to forthcoming
publications. 

\emph{Model \& methods.---} Ideal Josephson junctions with the conventional
harmonic current-phase relation $I=I_{c}\sin\phi$ shunted with a
simple circuit environment (parallel resistance $R$ and capacitance
$C$; see Fig~\ref{Fig_V-I}a) and current-biased by $I_{b}$ are
described by the Resistively and Capacitively Shunted Junction (RCSJ)
model \cite{McCumber68, *Stewart68}. The Langevin equations for the
phase difference $\phi(t)$ and voltage $V(t)=\frac{\hbar}{2e}\frac{d\phi(t)}{dt}$
across the junction are just the first Kirchhoff's law and the Josephson
voltage-phase relation reading in the dimensionless units \cite{Kautz96}

\begin{equation}
\begin{split}\frac{\partial v(\tau)}{\partial\tau} & =i_{b}-\gamma v(\tau)-\sin\phi(\tau)+\zeta(\tau),\\
v(\tau) & =\partial\phi(\tau)/\partial\tau.
\end{split}
\label{eq:Langevin-dimensionless}
\end{equation}
Dimensionless quantities are defined with help of the plasma frequency
$\omega_{p}=\sqrt{2eI_{c}/\hbar C}$ and the quality factor of the
circuit $Q=\omega_{p}RC\equiv\gamma^{-1}$ quantifying the level of
damping of the phase dynamics as: junction voltage $v=\tfrac{QV}{I_{c}R}$,
junction current $i=\tfrac{I}{I_{c}}$, time $\tau=\omega_{p}t$,
temperature $\Theta=\tfrac{2ek_{B}T}{\hbar I_{c}}$ (with the Boltzmann
constant $k_{B}$), bias current $i_{b}=\tfrac{I_{b}}{I_{c}}$, and
the Gaussian white noise $\zeta$ with the correlation functions $\langle\zeta(\tau)\rangle=0,\hspace{0.2cm}\langle\zeta(\tau_{1})\zeta(\tau_{2})\rangle=2\gamma\Theta\delta(\tau_{1}-\tau_{2})$.
Eqs.~\eqref{eq:Langevin-dimensionless} imply the associated Fokker-Planck
equation \cite{Risken} for the probability distribution function
$W(\phi,v,\tau)$ 
\begin{equation}
\begin{split}\dfrac{\partial}{\partial\tau}W(\phi,v;\tau) & =-v\dfrac{\partial}{\partial\phi}W+\dfrac{\partial}{\partial v}\left(\gamma v+\sin\phi-i_{b}+\gamma\Theta\dfrac{\partial}{\partial v}\right)W\\
 & \equiv L_{\mathrm{FP}}W(\phi,v;\tau).
\end{split}
\label{eq:Fokker-Planck}
\end{equation}

We are interested in the mean voltage $\langle v\rangle=\int\limits _{0}^{2\pi}\mathrm{d}\phi\int\limits _{-\infty}^{\infty}\mathrm{d}vvW_{\mathrm{stat}}(\phi,v),$
and the (zero-frequency) voltage noise $S=\int\limits _{-\infty}^{\infty}\mathrm{d}\tau\big(\langle v(\tau)v(0)\rangle-\langle v(\tau)\rangle\langle v(0)\rangle\big)$
in the stationary state $W_{\mathrm{stat}}(\phi,v)\equiv\lim_{\tau\to\infty}W(\phi,v;\tau)$
determined by the $2\pi$-periodic solution (in $\phi$) of the equation
$L_{\mathrm{FP}}W_{\mathrm{stat}}(\phi,v)=0$. Since the voltage autocorrelation
function is expressed as \cite[Sec.~7.2]{Risken} $\langle v(\tau)v(0)\rangle=\int\limits _{0}^{2\pi}\mathrm{d}\phi\int\limits _{-\infty}^{\infty}\mathrm{d}vve^{|\tau|L_{FP}}vW_{\mathrm{stat}}(\phi,v)$
the noise can be calculated by the formula $S=-2\int\limits _{0}^{2\pi}\mathrm{d}\phi\int\limits _{-\infty}^{\infty}\mathrm{d}vvR(\phi,v)$
via an auxiliary quantity $R(\phi,v)$ satisfying the equation $L_{\mathrm{FP}}R(\phi,v)=(v-\left\langle v\right\rangle )W_{\mathrm{stat}}(\phi,v)$
and conditions $R(\phi+2\pi,v)=R(\phi,v)$ and $\int\limits _{0}^{2\pi}\mathrm{d}\phi\int\limits _{-\infty}^{\infty}\mathrm{d}vR(\phi,v)=0$ \cite{SM}. We have found
both $W_{\mathrm{stat}}(\phi,v)$ and $R(\phi,v)$ numerically by
the MCF method \cite[Sec.~11.5]{Risken} which first expresses the
$v$-part of the equations in terms of quantum oscillator basis functions,
thus obtaining a tridiagonal coupled system of $\phi$-dependent differential
equations (Brinkmann hierarchy). The $2\pi$-periodic $\phi$-parts
are then expanded into the Fourier series and solved via the MCF as
explained in the Supplemental Material \cite{SM}. The method works
for an arbitrary current-phase relationship. Furthermore, using the
finite-frequency generalization of the problem exactly analogous to
Ref.~\cite{Flindt2005b}, we could easily evaluate also the finite-frequency
voltage noise. 

An alternative method for evaluation of the zero-frequency voltage
noise is to use the full counting statistics (FCS) approach pioneered
in this context in Ref.~\cite{Golubev10}. The aim of that method
is the calculation of the $k$-dependent ($k$ is the \emph{counting-field})
cumulant generating function (CGF) $F(k;\tau)\equiv\ln\int\limits _{-\infty}^{\infty}\mathrm{d}\phi e^{ik\phi}\intop\limits _{-\infty}^{\infty}dvW(\phi,v;\tau)$
from a non-stationary solution $W(\phi,v;\tau)$ of Eq.~\eqref{eq:Fokker-Planck}
\cite[Sec.~11.7]{Risken}. For long times $\tau\to\infty$ the CGF
generates \emph{all} stationary cumulants of the voltage by derivatives
with respect to $k$ at $k=0$ and its full $k$-dependence can be
used for evaluation of phase-slips rates as shown below. Following
the analogous derivations in Refs.~\cite{Bagrets03,Flindt05} we
obtain $\lim_{\tau\to\infty}F(k;\tau)/\tau=\lambda_{0}(k)$, where
$\lambda_{0}(k)$ is the counting-field-dependent eigenvalue of the
full problem \eqref{eq:Fokker-Planck} with modified boundary condition
$W_{0}(\phi+2\pi,v)=e^{-i2\pi k}W_{0}(\phi,v)$ \cite{Golubev10}
with the biggest real part (adiabatically developed with increasing
$k$ from the stationary solution $\lambda_{0}(k=0)=0$). This eigenvalue
can be also obtained by the MCF method as shown in the Supplemental
Material \cite{SM} and yields the mean voltage $\left\langle v\right\rangle \equiv-i\lambda_{0}^{'}(0)$,
voltage noise $S\equiv-\lambda_{0}^{''}(0)$, and similarly also the
higher-order voltage cumulants. We have verified that both calculation
methods give the same results for the mean voltage and noise. 

\begin{figure}
\centering{}\includegraphics[width=1\columnwidth]{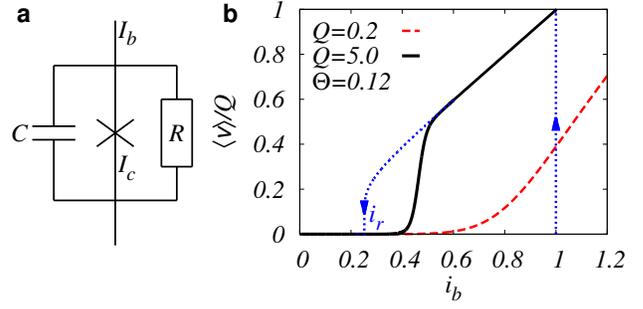}\caption{(Color online) a) Circuit representation of the RCSJ model. b) $\left\langle v\right\rangle -i_{b}$
characteristics for a weakly-damped junction $Q=5$ (black solid line)
and a strongly-damped one $Q=0.2$ (red dashed line). Blue dotted
lines represent the hysteretic $\left\langle v\right\rangle -i_{b}$
characteristics for forward and backward $i_{b}$ ramping of noiseless
case with $Q=5$. \label{Fig_V-I}}
 
\end{figure}

\emph{$\left\langle v\right\rangle -i_{b}$ characteristics.--- }In
Fig.~\ref{Fig_V-I}b we recapitulate for completeness the known results
\cite{Risken} for the mean voltage by plotting $\left\langle v\right\rangle -i_{b}$
curves for two complementary values of the quality factor representing
the strongly ($Q=0.2$) and weakly ($Q=5$) damped cases. While the
strong damping case shows with increasing $i_{b}$ a smooth crossover
from the nearly zero voltage (diffusive branch \cite{Kautz90}) to
the Ohmic behavior determined by $R$ for $i_{b}\gtrsim1$, the underdamped
curve exhibits a much sharper transition between the two regimes.
This can be understood as the noise-induced sudden switching between
two coexisting dynamical states of the underdamped junction in the
noiseless (zero temperature) limit represented by the dotted blue
lines in the figure revealing a strong hysteresis between the forward
and backward ramping of the bias current $i_{b}$. Using the well-known
mechanical analogy of the RCSJ model, which is the damped particle
in the tilted washboard potential $U(\phi)=\cos\phi+i_{b}\phi$ (illustrated
in the inset of Fig.~\ref{Fig_Fano-factor}a), one can easily see
that the zero voltage state (locked solution) is stable up to $i_{b}=1$
while a finite voltage state $v_{r}$ (running solution) is stable
down to the $Q$-dependent retrapping current $i_{r}(Q)$ determined
by the energy balance between the energy supply by the bias current
and dissipation \cite{Risken}. Without noise the originally trapped
``phase particle'' stays locked in the potential minimum until the
bias current is high enough to wipe off the local extremes of the
potential. On the other hand, if the particle is already running,
then, because of the inertia, it can still overcome the local maxima
and keep running if the damping is low enough. For finite temperatures
the stationary $\left\langle v\right\rangle -i_{b}$ characteristics
in the bistability region $i_{r}(Q)<i_{b}<1$ is the weighted average
of the running and locked solutions governed by the escape and retrapping
rates, which can be determined from the noise as demonstrated below. 

\begin{figure}
\centering{}\includegraphics[width=1\columnwidth]{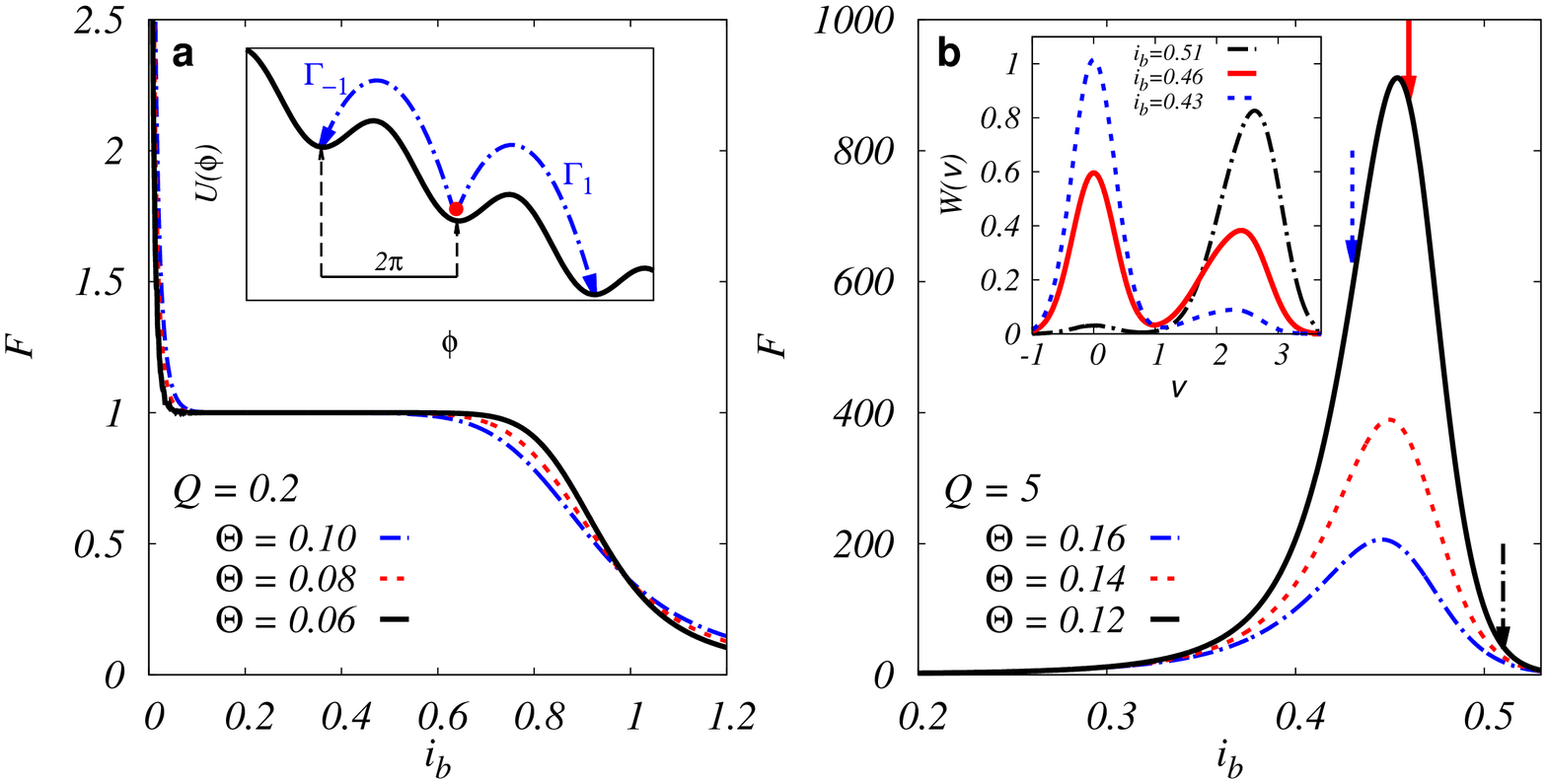}
\caption{(Color online) Fano factor for $Q=0.2$ (a) and $Q=5.0$ (b) and different
temperatures $\Theta$. Insets: mechanical analogue of Eq.~\eqref{eq:Langevin-dimensionless}
with $\Gamma_{1/-1}$ being the rates of the forward/backward single
phase-slips (a) and bimodal stationary distribution functions plotted
for bias current values marked with corresponding arrows (b). \label{Fig_Fano-factor}}
\end{figure}

\emph{Fano factor and switching process.---} Voltage noise properties
can be used to analyze the phase dynamics in far more detail. In Fig.~\ref{Fig_Fano-factor}
we plot the Fano factor $F\equiv S/2\pi v$ for strongly damped $Q=0.2$
(left panel a) and weakly damped $Q=5$ junctions (right panel b)
at different temperatures. Our numerical results for the strongly
damped case are very close to those for the overdamped RSJ model \cite{Golubev10}.
Because of low temperature $\Theta\ll1$ the low-bias-current behavior
for $i_{b}\lesssim0.6$ can be perfectly understood by the description
in terms of thermally-induced forward and backward single (i.e., by
$2\pi$) phase-slips shown in the inset of the left panel. This simple
picture yields for the Fano factor in this regime $F=\coth\pi i_{b}/\Theta$
\cite[Eq.~(28)]{Golubev10} exhibiting the characteristic divergence
at $i_{b}=0$ due to the finite thermal noise and the plateau at the
Poissonian value of $F=1$ for larger values of $i_{b}$. Above the
critical current $i_{b}>1$ the junction is in the running state and
the prevailing component of the noise is the simple Johnson thermal
noise of the resistor with $F=\Theta\tfrac{2i_{b}^{2}+1}{(i_{b}^{2}-1)^{\frac{3}{2}}}$
for $i_{b}\gtrsim1+\Theta^{2/3}$ \cite[Eqs.~(36),(37)]{Golubev10}.

\begin{figure}
\centering{}\includegraphics[width=1\columnwidth]{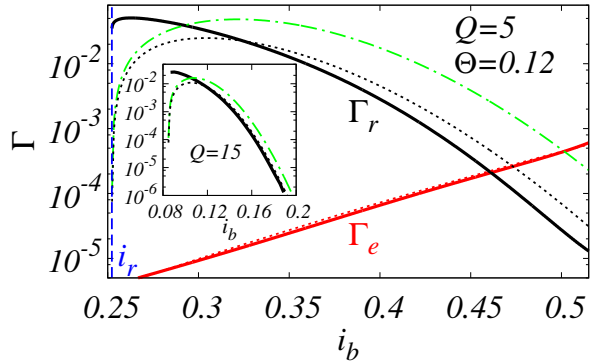}
\caption{(Color online) Escape rate $\Gamma_{e}$ and retrapping rate $\Gamma_{r}$
numerically computed from Eqs.~\eqref{E_GRL1} (solid lines) compared
with analytical approaches (dashed lines; for details see the main
text). Inset: More detailed comparison of the numerical retrapping
rate (full line) with Mel'nikov's (black dashed line) and Ben-Jacob's
formula (green dot-dashed line) for a junction with higher quality
factor $Q=15$. \label{Fig_rates}}
\end{figure}

The Fano factor vs.~bias current dependence of the underdamped circuit
in Fig.~\ref{Fig_Fano-factor}b is qualitatively different from the
strongly damped case. As shown in Fig.~\ref{Fig_phase-slips}a there
is also the low-$i_{b}$ thermal singularity in the Fano factor followed
for increasing $i_{b}\lesssim i_{r}(Q)$ by a flatter part due to
multiple phase-slips discussed in previous works \cite{Melnikov85,Kautz90,Melnikov:PhysRep91}
and studied in detail below. However, the dominant feature of the
plot in Fig.~\ref{Fig_Fano-factor}b is the strongly temperature-dependent
huge peak in the Fano factor ($F\approx100-1000$) around the switching
current (compare with Fig.~\ref{Fig_V-I}b) characteristic of the
dichotomous switching process. This interpretation is further supported
by the inset of Fig.~\ref{Fig_Fano-factor}b with the voltage distribution
function $W(v)\equiv\int_{0}^{2\pi}d\phi W_{\mathrm{stat}}(\phi,v)$
for various bias currents showing curves with well-separated double
peaks corresponding to differently weighted two metastable states.
The first peak centered around $v=0$ describes the locked phase while
the second one around the $i_{b}$-dependent noiseless running solution
$v_{r}$ reveals the running phase. Since $F\gg1$ around the peak
we can neglect noise contributions inherent to the metastable states
\cite{Jordan04} and use the average voltage $\left\langle v\right\rangle $
and Fano factor $F$ for the evaluation of the escape rate $\Gamma_{e}$
(from locked to running state) and retrapping rate $\Gamma_{r}$ (from
running to locked state) \cite{Flindt05} 
\begin{equation}
\Gamma_{e}=\frac{\left\langle v\right\rangle (v_{r}-\left\langle v\right\rangle )}{\pi v_{r}F},\hspace{0.3cm}\Gamma_{r}=\frac{(v_{r}-\left\langle v\right\rangle )^{2}}{\pi v_{r}F},\label{E_GRL1}
\end{equation}
which are presented in Fig.~\ref{Fig_rates} as solid lines and compared
with analytical predictions (dashed lines). On one hand, switching
from the locked phase to the running one happens just by thermally-induced
overcoming of the neighboring potential maximum and, therefore, $\Gamma_{e}=\tfrac{1}{2\pi}\left(\sqrt{\tfrac{\gamma^{2}}{4}+\sqrt{1-i_{b}^{2}}}-\tfrac{\gamma}{2}\right)e^{-\tfrac{2(i_{b}\arcsin i_{b}+\sqrt{1-i_{b}^{2}})-\pi i_{b}}{\Theta}}$
is given by the Kramers escape rate from a potential well \cite{Hanggi90}.
One can see that this prediction is in an excellent agreement with
our numerical result. On the other hand, the retrapping problem is
far more complicated and has not been addressed in the whole parameter
regime, only asymptotic solutions in various limits exist. 

Two mutually inconsistent analytical approaches by Ben-Jacob et al.~\cite{BenJacob82}
and Mel'nikov \cite{Melnikov87,Melnikov:PhysRep91} were introduced
for the limit of very weak damping $Q\to\infty$. In Fig.~\ref{Fig_rates}
we compare our results with those predictions and find a rather poor
agreement with either one. With increasing $Q$ Melnikov's approach
seems to asymptotically approach our results for sufficiently large
$i_{b}$ as shown in the inset. Ben-Jacob's prediction on the other
hand remains off which might be connected with its existing critiques
\cite{Jung83,Melnikov87}. The origin of the obvious and persistent
discrepancy between Melnikov's and our approach for $i_{b}$ close
to the onset of the bistability region is unclear, but we suspect
it is connected with the very existence, stability, and definition
of the running state for the noisy case close to the bifurcation point.
In any case, the existing theories perform quite badly for the moderately
damped junctions while our numerical method allows to extract the
retrapping rate fast and reliably in a wide range of junction parameters,
in particular for arbitrary values of the quality factor $Q$ and
bias current $i_{b}$ which is the crucial ingredient for the description
of switching experiments such as the recent ones in Ref.~\cite{Lukens:PRB05, *Delsing:PRL05, *Pekola:PRL05, *Delsing:PRB07, *Levchenko:PRL13}. 

\emph{Phase-slips.---} Now we turn our attention to the MPS regime
of bias current smaller than the onset of the switching regime $i_{b}\lesssim i_{r}(Q)$
shown Fig.~\eqref{Fig_phase-slips}a. Analogously to the overdamped
case \cite{Golubev10,Challis:PRE13} the solution to the Fokker-Planck
equation \eqref{eq:Fokker-Planck} in this regime can be approximated
by a weighted sum of quasi-equilibrated sharp ($\Theta\ll1$) Gaussian
distributions around the local minima $W(\phi,v;\tau)\approx\sum_{m}P_{m}(\tau)w(\phi-2\pi m,v)$
with $w(\phi,v)=\tfrac{\sqrt[4]{1-i_{b}^{2}}\exp\left(-(\phi-\arcsin i_{b})^{2}\sqrt{1-i_{b}^{2}}/2\Theta\right)\exp(-v^{2}/2\Theta)}{2\pi\Theta}$
\cite[Eq.~(22)]{Golubev10} and time-dependent weights $P_{m}(\tau)$.
These are \emph{assumed} to satisfy the (Markovian) master equation
\begin{equation}
\begin{aligned}\frac{dP_{m}(\tau)}{d\tau} & =\sum_{n\neq0}\Gamma_{n}P_{m-n}(\tau)-\sum_{n\neq0}\Gamma_{n}P_{m}(\tau)\\
 & \equiv\sum_{n}\Gamma_{n}P_{m-n}(\tau),\text{ {with} }\Gamma_{0}\equiv-\sum_{n\neq0}\Gamma_{n}.
\end{aligned}
\label{eq:ME1}
\end{equation}
Here $\Gamma_{n}\ (n\neq0)$ are the rates of elementary MPS by $2\pi n$
(negative $n$ correspond to backward rates against the bias). 

To find the MPS rates we use the FCS methodology introduced in Ref.~\cite{Golubev10}
for the RSJ model combined with the procedure of identification of
elementary processes \cite{Vanevic:PRL07, *Vanevic:PRB08, *Padurariu:PRB12}.
If the master equation \eqref{eq:ME1} description is a good approximation
of the full phase dynamics we can evaluate the CGF from it and equate
that with the full CGF calculated by the MCF. We have for the approximate
probability density $\exp\left[F(k;\tau)\right]\approx\sum_{m}P_{m}(\tau)e^{2\pi ikm}\tilde{w}(k)\equiv\mathcal{P}(k;\tau)\cdot\tilde{w}(k)$,
with $\tilde{w}(k)=\exp\left(ik\arcsin i_{b}-2\Theta k^{2}/\sqrt{1-i_{b}^{2}}\right)$
and $\mathcal{P}(k;\tau)\equiv\sum_{m}P_{m}(\tau)e^{2\pi ikm}$ satisfying
the $k$-dependent differential equation $\frac{d\mathcal{P}(k;\tau)}{d\tau}=\left(\sum_{n}\Gamma_{n}e^{2\pi ikn}-\sum_{n}\Gamma_{n}\right)\mathcal{P}(k;\tau)$.
This allows us to identify $\lambda_{0}(k)=\sum_{n}\Gamma_{n}(e^{2\pi ikn}-1)$
in the MPS regime describing a mixture of independent Poissonian processes
of phase-slips by $2\pi n$ whose rates can be evaluated as $\Gamma_{n}=\int\limits _{0}^{1}\mathrm{d}k\lambda_{0}(k)e^{-2\pi ikn}$
for $n\neq0$. Importantly, the method itself provides tools for checking
its validity by comparing the approximate mean voltage $\left\langle v\right\rangle =2\pi\sum_{n}\Gamma_{n}n$
and Fano factor $F=\sum_{n}\Gamma_{n}n^{2}/\sum_{n}\Gamma_{n}n$ with
those computed directly by the MCF method. In Fig.~\ref{Fig_phase-slips}a
it is explicitly shown that the two Fano-factors perfectly match up
to the values of $i_{b}$ where the switching process sets in (and
correspondence eventually breaks down). 
\begin{figure}
\includegraphics[width=1\columnwidth]{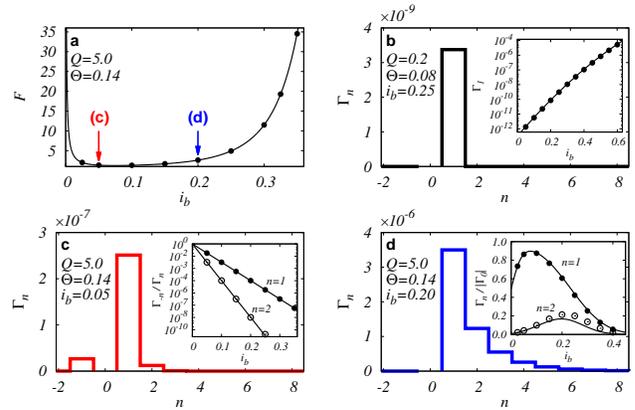} \caption{(Color online) a) Detail of the $\Theta=0.14$ Fano factor plot from
Fig.~\ref{Fig_Fano-factor}b in the small-$i_{b}$ regime where multiple
phase-slips are the prevailing source of voltage noise. Full line
is the Fano factor calculated from the full Fokker-Planck equation
\eqref{eq:Fokker-Planck} while the overlapping dots are the checks
evaluated via the MPS rates (see the main text). b) Single-phase-slip
rate for a strongly damped junction with $Q=0.2$. Inset: comparison
of the $i_{b}$-dependence of the rate evaluated numerically (dots)
and by the Kramers formula for the overdamped case (solid line). c)
and d) Rates of MPS (of order $n$) for two values of the bias current
$i_{b}$ denoted in panel a) by the corresponding arrows. Insets:
verification of the detailed balance condition (c) and comparison
of the $i_{b}$-dependence of the first two normalized phase-slip
rates (respective dots) with the Mel'nikov formula (lines) (d).\label{Fig_phase-slips}}
\end{figure}

In Fig.~\ref{Fig_phase-slips} we plot MPS rates $\Gamma_{n}$ for
strong-damping case $Q=0.2$ (b) and weak damping case $Q=5;\: i_{b}=0.05,\,0.20$
(c, d). Only the single phase-slips are realized in the strong-damping
case, which is consistent with the Fano-factor plateau at one in Fig.~\ref{Fig_Fano-factor}a.
The dependence of $\Gamma_{1}$ on $i_{b}$ in the regime of the plateau
is plotted in the inset with dots and compared with the Kramers formula
for escape across the adjacent barrier \cite{Kramers40,Golubev10,Hanggi90}
(solid line in the inset). The behavior of the weak-damping case is
much richer. The presence of multiple (double) phase-slips (alongside
the single backward phase-slips) is evident even for small bias currents
in Fig.~\ref{Fig_phase-slips}c and their importance increases with
increasing bias current (Fig.~\ref{Fig_phase-slips}d). The inset
depicts the normalized (by $|\Gamma_{0}|=\sum_{n\neq0}\Gamma_{n}$)
rates of the single and double phase-slips as functions of $i_{b}$
(dots) together with Mel'nikov's approximative asymptotic formulas
valid for $Q\to\infty$ \cite{Melnikov85,Melnikov:PhysRep91} (solid
lines in the inset). We can see reasonable agreement which analogously
to Fig.~\ref{Fig_rates} further improves with increasing $Q$. In
the inset of Fig.~\ref{Fig_phase-slips}c we test that the ratio
$\Gamma_{-n}/\Gamma_{n}=\exp(-2\pi ni_{b}/\Theta)$ for $n=1,\,2$
satisfies the detailed balance condition with the potential drop $2\pi i_{b}n$
along the phase-slips of the given order $n$. Altogether, our method
provides reliable results for the MPS rates for a wide class of junctions. 

\emph{Conclusion.--- }To summarize, we have developed an efficient
and reliable numerical scheme based on the matrix continued fraction
for the analysis of phase dynamics in arbitrarily damped current-biased
Josephson junctions. It allows us to study the voltage noise and,
consequently, analyze in detail various transport regimes, in particular
the switching regime providing us with the escape and retrapping rates
and the regime of multiple phase-slips with their full characterization
in terms of the corresponding rates. The method can be generalized
from the RCSJ model to other experimentally relevant cases with modified
circuits and/or more complicated microscopic current-phase relations
as well as extended to the evaluation of the frequency-dependent voltage
noise. 

\emph{Acknowledgments.--- }We thank Pertti Hakonen, Tero Heikkil\"{a},
and Gabriel Niebler for stimulating discussions drawing our attention
to this problem. We acknowledge financial support by the Czech Science
Foundation via Grant No.~204/11/J042 (M.\v{Z}. and T.N.) and by the DFG
via BE 3803/5 (W.B.).

\bibliographystyle{apsrev4-1}

\begin{thebibliography}{41}%
\makeatletter
\providecommand \@ifxundefined [1]{%
 \@ifx{#1\undefined}
}%
\providecommand \@ifnum [1]{%
 \ifnum #1\expandafter \@firstoftwo
 \else \expandafter \@secondoftwo
 \fi
}%
\providecommand \@ifx [1]{%
 \ifx #1\expandafter \@firstoftwo
 \else \expandafter \@secondoftwo
 \fi
}%
\providecommand \natexlab [1]{#1}%
\providecommand \enquote  [1]{``#1''}%
\providecommand \bibnamefont  [1]{#1}%
\providecommand \bibfnamefont [1]{#1}%
\providecommand \citenamefont [1]{#1}%
\providecommand \href@noop [0]{\@secondoftwo}%
\providecommand \href [0]{\begingroup \@sanitize@url \@href}%
\providecommand \@href[1]{\@@startlink{#1}\@@href}%
\providecommand \@@href[1]{\endgroup#1\@@endlink}%
\providecommand \@sanitize@url [0]{\catcode `\\12\catcode `\$12\catcode
  `\&12\catcode `\#12\catcode `\^12\catcode `\_12\catcode `\%12\relax}%
\providecommand \@@startlink[1]{}%
\providecommand \@@endlink[0]{}%
\providecommand \url  [0]{\begingroup\@sanitize@url \@url }%
\providecommand \@url [1]{\endgroup\@href {#1}{\urlprefix }}%
\providecommand \urlprefix  [0]{URL }%
\providecommand \Eprint [0]{\href }%
\providecommand \doibase [0]{http://dx.doi.org/}%
\providecommand \selectlanguage [0]{\@gobble}%
\providecommand \bibinfo  [0]{\@secondoftwo}%
\providecommand \bibfield  [0]{\@secondoftwo}%
\providecommand \translation [1]{[#1]}%
\providecommand \BibitemOpen [0]{}%
\providecommand \bibitemStop [0]{}%
\providecommand \bibitemNoStop [0]{.\EOS\space}%
\providecommand \EOS [0]{\spacefactor3000\relax}%
\providecommand \BibitemShut  [1]{\csname bibitem#1\endcsname}%
\let\auto@bib@innerbib\@empty
\bibitem [{\citenamefont {{Barone}}\ and\ \citenamefont
  {{Patern\`{o}}}(1982)}]{Barone}%
  \BibitemOpen
  \bibfield  {author} {\bibinfo {author} {\bibfnamefont {A.}~\bibnamefont
  {{Barone}}}\ and\ \bibinfo {author} {\bibfnamefont {G.}~\bibnamefont
  {{Patern\`{o}}}},\ }\href@noop {} {\emph {\bibinfo {title} {Physics and
  Applications of the Josephson Effect}}}\ (\bibinfo  {publisher} {A
  Willey-Interscience Publications, John Wiley \& Sons, New York, Chichester,
  Brisbane, Toronto, Singapore},\ \bibinfo {year} {1982})\BibitemShut {NoStop}%
\bibitem [{\citenamefont {{Likharev}}(1986)}]{Likharev86}%
  \BibitemOpen
  \bibfield  {author} {\bibinfo {author} {\bibfnamefont {K.~K.}\ \bibnamefont
  {{Likharev}}},\ }\href@noop {} {\emph {\bibinfo {title} {Dynamics of
  Josephson Junctions and Circuits}}}\ (\bibinfo  {publisher} {Gordon and
  Breach, New York},\ \bibinfo {year} {1986})\BibitemShut {NoStop}%
\bibitem [{\citenamefont {{Kautz}}(1996)}]{Kautz96}%
  \BibitemOpen
  \bibfield  {author} {\bibinfo {author} {\bibfnamefont {R.~L.}\ \bibnamefont
  {{Kautz}}},\ }\href {\doibase 10.1088/0034-4885/59/8/001} {\bibfield
  {journal} {\bibinfo  {journal} {Reports on Progress in Physics}\ }\textbf
  {\bibinfo {volume} {59}},\ \bibinfo {pages} {935} (\bibinfo {year}
  {1996})}\BibitemShut {NoStop}%
\bibitem [{\citenamefont {{Risken}}(1989)}]{Risken}%
  \BibitemOpen
  \bibfield  {author} {\bibinfo {author} {\bibfnamefont {H.}~\bibnamefont
  {{Risken}}},\ }\href@noop {} {\emph {\bibinfo {title} {The Fokker-Planck
  Equation, 2nd ed.}}}\ (\bibinfo  {publisher} {Springer-Verlag, Berlin,
  Heidelberg},\ \bibinfo {year} {1989})\BibitemShut {NoStop}%
\bibitem [{\citenamefont {Buzdin}(2005)}]{Budzin05}%
  \BibitemOpen
  \bibfield  {author} {\bibinfo {author} {\bibfnamefont {A.~I.}\ \bibnamefont
  {Buzdin}},\ }\href {\doibase 10.1103/RevModPhys.77.935} {\bibfield  {journal}
  {\bibinfo  {journal} {Rev. Mod. Phys.}\ }\textbf {\bibinfo {volume} {77}},\
  \bibinfo {pages} {935} (\bibinfo {year} {2005})}\BibitemShut {NoStop}%
\bibitem [{\citenamefont {Sickinger}\ \emph {et~al.}(2012)\citenamefont
  {Sickinger}, \citenamefont {Lipman}, \citenamefont {Weides}, \citenamefont
  {Mints}, \citenamefont {Kohlstedt}, \citenamefont {Koelle}, \citenamefont
  {Kleiner},\ and\ \citenamefont {Goldobin}}]{Sickinger12}%
  \BibitemOpen
  \bibfield  {author} {\bibinfo {author} {\bibfnamefont {H.}~\bibnamefont
  {Sickinger}}, \bibinfo {author} {\bibfnamefont {A.}~\bibnamefont {Lipman}},
  \bibinfo {author} {\bibfnamefont {M.}~\bibnamefont {Weides}}, \bibinfo
  {author} {\bibfnamefont {R.~G.}\ \bibnamefont {Mints}}, \bibinfo {author}
  {\bibfnamefont {H.}~\bibnamefont {Kohlstedt}}, \bibinfo {author}
  {\bibfnamefont {D.}~\bibnamefont {Koelle}}, \bibinfo {author} {\bibfnamefont
  {R.}~\bibnamefont {Kleiner}}, \ and\ \bibinfo {author} {\bibfnamefont
  {E.}~\bibnamefont {Goldobin}},\ }\href {\doibase
  10.1103/PhysRevLett.109.107002} {\bibfield  {journal} {\bibinfo  {journal}
  {Phys. Rev. Lett.}\ }\textbf {\bibinfo {volume} {109}},\ \bibinfo {pages}
  {107002} (\bibinfo {year} {2012})}\BibitemShut {NoStop}%
\bibitem [{\citenamefont {De~Franceschi}\ \emph {et~al.}(2010)\citenamefont
  {De~Franceschi}, \citenamefont {Kouwenhoven}, \citenamefont
  {Sch{\"o}nenberger},\ and\ \citenamefont {Wernsdorfer}}]{De-Franceschi10}%
  \BibitemOpen
  \bibfield  {author} {\bibinfo {author} {\bibfnamefont {S.}~\bibnamefont
  {De~Franceschi}}, \bibinfo {author} {\bibfnamefont {L.}~\bibnamefont
  {Kouwenhoven}}, \bibinfo {author} {\bibfnamefont {C.}~\bibnamefont
  {Sch{\"o}nenberger}}, \ and\ \bibinfo {author} {\bibfnamefont
  {W.}~\bibnamefont {Wernsdorfer}},\ }\href
  {http://dx.doi.org/10.1038/nnano.2010.173} {\bibfield  {journal} {\bibinfo
  {journal} {Nat Nano}\ }\textbf {\bibinfo {volume} {5}},\ \bibinfo {pages}
  {703} (\bibinfo {year} {2010})}\BibitemShut {NoStop}%
\bibitem [{\citenamefont {Goldobin}\ \emph {et~al.}(2013)\citenamefont
  {Goldobin}, \citenamefont {Kleiner}, \citenamefont {Koelle},\ and\
  \citenamefont {Mints}}]{Goldobin:PRL13}%
  \BibitemOpen
  \bibfield  {author} {\bibinfo {author} {\bibfnamefont {E.}~\bibnamefont
  {Goldobin}}, \bibinfo {author} {\bibfnamefont {R.}~\bibnamefont {Kleiner}},
  \bibinfo {author} {\bibfnamefont {D.}~\bibnamefont {Koelle}}, \ and\ \bibinfo
  {author} {\bibfnamefont {R.~G.}\ \bibnamefont {Mints}},\ }\href
  {http://link.aps.org/doi/10.1103/PhysRevLett.111.057004} {\bibfield
  {journal} {\bibinfo  {journal} {Physical Review Letters}\ }\textbf {\bibinfo
  {volume} {111}},\ \bibinfo {pages} {057004} (\bibinfo {year}
  {2013})}\BibitemShut {NoStop}%
\bibitem [{\citenamefont {Martinis}\ and\ \citenamefont
  {Kautz}(1989)}]{Kautz89}%
  \BibitemOpen
  \bibfield  {author} {\bibinfo {author} {\bibfnamefont {J.~M.}\ \bibnamefont
  {Martinis}}\ and\ \bibinfo {author} {\bibfnamefont {R.~L.}\ \bibnamefont
  {Kautz}},\ }\href {\doibase 10.1103/PhysRevLett.63.1507} {\bibfield
  {journal} {\bibinfo  {journal} {Phys. Rev. Lett.}\ }\textbf {\bibinfo
  {volume} {63}},\ \bibinfo {pages} {1507} (\bibinfo {year}
  {1989})}\BibitemShut {NoStop}%
\bibitem [{\citenamefont {Kautz}\ and\ \citenamefont
  {Martinis}(1990)}]{Kautz90}%
  \BibitemOpen
  \bibfield  {author} {\bibinfo {author} {\bibfnamefont {R.~L.}\ \bibnamefont
  {Kautz}}\ and\ \bibinfo {author} {\bibfnamefont {J.~M.}\ \bibnamefont
  {Martinis}},\ }\href {\doibase 10.1103/PhysRevB.42.9903} {\bibfield
  {journal} {\bibinfo  {journal} {Phys. Rev. B}\ }\textbf {\bibinfo {volume}
  {42}},\ \bibinfo {pages} {9903} (\bibinfo {year} {1990})}\BibitemShut
  {NoStop}%
\bibitem [{\citenamefont {Fenton}\ and\ \citenamefont
  {Warburton}(2008)}]{Fenton08}%
  \BibitemOpen
  \bibfield  {author} {\bibinfo {author} {\bibfnamefont {J.~C.}\ \bibnamefont
  {Fenton}}\ and\ \bibinfo {author} {\bibfnamefont {P.~A.}\ \bibnamefont
  {Warburton}},\ }\href {\doibase 10.1103/PhysRevB.78.054526} {\bibfield
  {journal} {\bibinfo  {journal} {Phys. Rev. B}\ }\textbf {\bibinfo {volume}
  {78}},\ \bibinfo {pages} {054526} (\bibinfo {year} {2008})}\BibitemShut
  {NoStop}%
\bibitem [{\citenamefont {{Voss}}(1981)}]{Voss81}%
  \BibitemOpen
  \bibfield  {author} {\bibinfo {author} {\bibfnamefont {R.~F.}\ \bibnamefont
  {{Voss}}},\ }\href {\doibase 10.1007/BF00116701} {\bibfield  {journal}
  {\bibinfo  {journal} {Journal of Low Temperature Physics}\ }\textbf {\bibinfo
  {volume} {42}},\ \bibinfo {pages} {151} (\bibinfo {year} {1981})}\BibitemShut
  {NoStop}%
\bibitem [{\citenamefont {{Golubev}}\ \emph {et~al.}(2010)\citenamefont
  {{Golubev}}, \citenamefont {{Marthaler}}, \citenamefont {{Utsumi}},\ and\
  \citenamefont {{Sch\"on}}}]{Golubev10}%
  \BibitemOpen
  \bibfield  {author} {\bibinfo {author} {\bibfnamefont {D.~S.}\ \bibnamefont
  {{Golubev}}}, \bibinfo {author} {\bibfnamefont {M.}~\bibnamefont
  {{Marthaler}}}, \bibinfo {author} {\bibfnamefont {Y.}~\bibnamefont
  {{Utsumi}}}, \ and\ \bibinfo {author} {\bibfnamefont {G.}~\bibnamefont
  {{Sch\"on}}},\ }\href {\doibase 10.1103/PhysRevB.81.184516} {\bibfield
  {journal} {\bibinfo  {journal} {Phys. Rev. B}\ }\textbf {\bibinfo {volume}
  {81}},\ \bibinfo {pages} {184516} (\bibinfo {year} {2010})}\BibitemShut
  {NoStop}%
\bibitem [{\citenamefont {{\v Z}onda}\ and\ \citenamefont
  {Novotn{\'y}}(2012)}]{Zonda:PhScr12}%
  \BibitemOpen
  \bibfield  {author} {\bibinfo {author} {\bibfnamefont {M.}~\bibnamefont {{\v
  Z}onda}}\ and\ \bibinfo {author} {\bibfnamefont {T.}~\bibnamefont
  {Novotn{\'y}}},\ }\href
  {http://stacks.iop.org/1402-4896/2012/i=T151/a=014023} {\bibfield  {journal}
  {\bibinfo  {journal} {Physica Scripta}\ }\textbf {\bibinfo {volume} {2012}},\
  \bibinfo {pages} {014023} (\bibinfo {year} {2012})}\BibitemShut {NoStop}%
\bibitem [{Note1()}]{Note1}%
  \BibitemOpen
  \bibinfo {note} {P. Hakonen, private communication.}\BibitemShut {Stop}%
\bibitem [{\citenamefont {Vanevi{\'c}}\ \emph {et~al.}(2007)\citenamefont
  {Vanevi{\'c}}, \citenamefont {Nazarov},\ and\ \citenamefont
  {Belzig}}]{Vanevic:PRL07}%
  \BibitemOpen
  \bibfield  {author} {\bibinfo {author} {\bibfnamefont {M.}~\bibnamefont
  {Vanevi{\'c}}}, \bibinfo {author} {\bibfnamefont {Y.~V.}\ \bibnamefont
  {Nazarov}}, \ and\ \bibinfo {author} {\bibfnamefont {W.}~\bibnamefont
  {Belzig}},\ }\href {http://link.aps.org/doi/10.1103/PhysRevLett.99.076601}
  {\bibfield  {journal} {\bibinfo  {journal} {Physical Review Letters}\
  }\textbf {\bibinfo {volume} {99}},\ \bibinfo {pages} {076601} (\bibinfo
  {year} {2007})}\BibitemShut {NoStop}%
\bibitem [{\citenamefont {Vanevi{\'c}}\ \emph {et~al.}(2008)\citenamefont
  {Vanevi{\'c}}, \citenamefont {Nazarov},\ and\ \citenamefont
  {Belzig}}]{Vanevic:PRB08}%
  \BibitemOpen
  \bibfield  {author} {\bibinfo {author} {\bibfnamefont {M.}~\bibnamefont
  {Vanevi{\'c}}}, \bibinfo {author} {\bibfnamefont {Y.~V.}\ \bibnamefont
  {Nazarov}}, \ and\ \bibinfo {author} {\bibfnamefont {W.}~\bibnamefont
  {Belzig}},\ }\href {http://link.aps.org/doi/10.1103/PhysRevB.78.245308}
  {\bibfield  {journal} {\bibinfo  {journal} {Physical Review B}\ }\textbf
  {\bibinfo {volume} {78}},\ \bibinfo {pages} {245308} (\bibinfo {year}
  {2008})}\BibitemShut {NoStop}%
\bibitem [{\citenamefont {Padurariu}\ \emph {et~al.}(2012)\citenamefont
  {Padurariu}, \citenamefont {Hassler},\ and\ \citenamefont
  {Nazarov}}]{Padurariu:PRB12}%
  \BibitemOpen
  \bibfield  {author} {\bibinfo {author} {\bibfnamefont {C.}~\bibnamefont
  {Padurariu}}, \bibinfo {author} {\bibfnamefont {F.}~\bibnamefont {Hassler}},
  \ and\ \bibinfo {author} {\bibfnamefont {Y.~V.}\ \bibnamefont {Nazarov}},\
  }\href {http://link.aps.org/doi/10.1103/PhysRevB.86.054514} {\bibfield
  {journal} {\bibinfo  {journal} {Physical Review B}\ }\textbf {\bibinfo
  {volume} {86}},\ \bibinfo {pages} {054514} (\bibinfo {year}
  {2012})}\BibitemShut {NoStop}%
\bibitem [{\citenamefont {Joyez}\ \emph {et~al.}(1999)\citenamefont {Joyez},
  \citenamefont {Vion}, \citenamefont {G{\"o}tz}, \citenamefont {Devoret},\
  and\ \citenamefont {Esteve}}]{Joyez99}%
  \BibitemOpen
  \bibfield  {author} {\bibinfo {author} {\bibfnamefont {P.}~\bibnamefont
  {Joyez}}, \bibinfo {author} {\bibfnamefont {D.}~\bibnamefont {Vion}},
  \bibinfo {author} {\bibfnamefont {M.}~\bibnamefont {G{\"o}tz}}, \bibinfo
  {author} {\bibfnamefont {M.}~\bibnamefont {Devoret}}, \ and\ \bibinfo
  {author} {\bibfnamefont {D.}~\bibnamefont {Esteve}},\ }\href {\doibase
  10.1023/A:1007733009637} {\bibfield  {journal} {\bibinfo  {journal} {Journal
  of Superconductivity}\ }\textbf {\bibinfo {volume} {12}},\ \bibinfo {pages}
  {757} (\bibinfo {year} {1999})}\BibitemShut {NoStop}%
\bibitem [{\citenamefont {Webster}\ \emph {et~al.}(2013)\citenamefont
  {Webster}, \citenamefont {Fenton}, \citenamefont {Hongisto}, \citenamefont
  {Giblin}, \citenamefont {Zorin},\ and\ \citenamefont
  {Warburton}}]{Webster13}%
  \BibitemOpen
  \bibfield  {author} {\bibinfo {author} {\bibfnamefont {C.~H.}\ \bibnamefont
  {Webster}}, \bibinfo {author} {\bibfnamefont {J.~C.}\ \bibnamefont {Fenton}},
  \bibinfo {author} {\bibfnamefont {T.~T.}\ \bibnamefont {Hongisto}}, \bibinfo
  {author} {\bibfnamefont {S.~P.}\ \bibnamefont {Giblin}}, \bibinfo {author}
  {\bibfnamefont {A.~B.}\ \bibnamefont {Zorin}}, \ and\ \bibinfo {author}
  {\bibfnamefont {P.~A.}\ \bibnamefont {Warburton}},\ }\href {\doibase
  10.1103/PhysRevB.87.144510} {\bibfield  {journal} {\bibinfo  {journal} {Phys.
  Rev. B}\ }\textbf {\bibinfo {volume} {87}},\ \bibinfo {pages} {144510}
  (\bibinfo {year} {2013})}\BibitemShut {NoStop}%
\bibitem [{\citenamefont {{McCumber}}(1968)}]{McCumber68}%
  \BibitemOpen
  \bibfield  {author} {\bibinfo {author} {\bibfnamefont {D.~E.}\ \bibnamefont
  {{McCumber}}},\ }\href {\doibase 10.1063/1.1656743} {\bibfield  {journal}
  {\bibinfo  {journal} {Journal of Applied Physics}\ }\textbf {\bibinfo
  {volume} {39}},\ \bibinfo {pages} {3113} (\bibinfo {year}
  {1968})}\BibitemShut {NoStop}%
\bibitem [{\citenamefont {Stewart}(1968)}]{Stewart68}%
  \BibitemOpen
  \bibfield  {author} {\bibinfo {author} {\bibfnamefont {W.~C.}\ \bibnamefont
  {Stewart}},\ }\href {http://dx.doi.org/10.1063/1.1651991} {\bibfield
  {journal} {\bibinfo  {journal} {Applied Physics Letters}\ }\textbf {\bibinfo
  {volume} {12}},\ \bibinfo {pages} {277} (\bibinfo {year} {1968})}\BibitemShut
  {NoStop}%
\bibitem [{\citenamefont {Flindt}\ \emph {et~al.}(2004)\citenamefont {Flindt},
  \citenamefont {Novotn\'{y}},\ and\ \citenamefont {Jauho}}]{Flindt2004}%
  \BibitemOpen
  \bibfield  {author} {\bibinfo {author} {\bibfnamefont {C.}~\bibnamefont
  {Flindt}}, \bibinfo {author} {\bibfnamefont {T.}~\bibnamefont {Novotn\'{y}}},
  \ and\ \bibinfo {author} {\bibfnamefont {A.-P.}\ \bibnamefont {Jauho}},\
  }\href@noop {} {\bibfield  {journal} {\bibinfo  {journal} {Phys. Rev. B}\
  }\textbf {\bibinfo {volume} {70}},\ \bibinfo {pages} {205334} (\bibinfo
  {year} {2004})}\BibitemShut {NoStop}%
\bibitem [{\citenamefont {{\v Z}onda}\ \emph {et~al.}()\citenamefont {{\v
  Z}onda}, \citenamefont {Belzig},\ and\ \citenamefont {Novotn{\'y}}}]{SM}%
  \BibitemOpen
  \bibfield  {author} {\bibinfo {author} {\bibfnamefont {M.}~\bibnamefont {{\v
  Z}onda}}, \bibinfo {author} {\bibfnamefont {W.}~\bibnamefont {Belzig}}, \
  and\ \bibinfo {author} {\bibfnamefont {T.}~\bibnamefont {Novotn{\'y}}},\
  }\href@noop {} {\enquote {\bibinfo {title} {Supplemental material}}\ }\BibitemShut {NoStop}%
\bibitem [{\citenamefont {Flindt}\ \emph {et~al.}(2005)\citenamefont {Flindt},
  \citenamefont {Novotn\'{y}},\ and\ \citenamefont {Jauho}}]{Flindt2005b}%
  \BibitemOpen
  \bibfield  {author} {\bibinfo {author} {\bibfnamefont {C.}~\bibnamefont
  {Flindt}}, \bibinfo {author} {\bibfnamefont {T.}~\bibnamefont {Novotn\'{y}}},
  \ and\ \bibinfo {author} {\bibfnamefont {A.-P.}\ \bibnamefont {Jauho}},\
  }\href@noop {} {\bibfield  {journal} {\bibinfo  {journal} {Physica E}\
  }\textbf {\bibinfo {volume} {29}},\ \bibinfo {pages} {411} (\bibinfo {year}
  {2005})}\BibitemShut {NoStop}%
\bibitem [{\citenamefont {Bagrets}\ and\ \citenamefont
  {Nazarov}(2003)}]{Bagrets03}%
  \BibitemOpen
  \bibfield  {author} {\bibinfo {author} {\bibfnamefont {D.~A.}\ \bibnamefont
  {Bagrets}}\ and\ \bibinfo {author} {\bibfnamefont {Y.~V.}\ \bibnamefont
  {Nazarov}},\ }\href {\doibase 10.1103/PhysRevB.67.085316} {\bibfield
  {journal} {\bibinfo  {journal} {Phys. Rev. B}\ }\textbf {\bibinfo {volume}
  {67}},\ \bibinfo {pages} {085316} (\bibinfo {year} {2003})}\BibitemShut
  {NoStop}%
\bibitem [{\citenamefont {{Flindt}}\ \emph {et~al.}(2005)\citenamefont
  {{Flindt}}, \citenamefont {{Novotn{\'y}}},\ and\ \citenamefont
  {{Jauho}}}]{Flindt05}%
  \BibitemOpen
  \bibfield  {author} {\bibinfo {author} {\bibfnamefont {C.}~\bibnamefont
  {{Flindt}}}, \bibinfo {author} {\bibfnamefont {T.}~\bibnamefont
  {{Novotn{\'y}}}}, \ and\ \bibinfo {author} {\bibfnamefont {A.-P.}\
  \bibnamefont {{Jauho}}},\ }\href {\doibase 10.1209/epl/i2004-10351-x}
  {\bibfield  {journal} {\bibinfo  {journal} {Europhys. Lett.}\ }\textbf
  {\bibinfo {volume} {69}},\ \bibinfo {pages} {475} (\bibinfo {year} {2005})} \BibitemShut {NoStop}%
\bibitem [{\citenamefont {{Mel'nikov}}(1985)}]{Melnikov85}%
  \BibitemOpen
  \bibfield  {author} {\bibinfo {author} {\bibfnamefont {V.~I.}\ \bibnamefont
  {{Mel'nikov}}},\ }\href@noop {} {\bibfield  {journal} {\bibinfo  {journal}
  {Zh. Eksp. Teor. Fiz.}\ }\textbf {\bibinfo {volume} {18}},\ \bibinfo {pages}
  {1429} (\bibinfo {year} {1985})}\BibitemShut {NoStop}%
\bibitem [{\citenamefont {Mel'nikov}(1991)}]{Melnikov:PhysRep91}%
  \BibitemOpen
  \bibfield  {author} {\bibinfo {author} {\bibfnamefont {V.~I.}\ \bibnamefont
  {Mel'nikov}},\ }\href {\doibase
  http://dx.doi.org/10.1016/0370-1573(91)90108-X} {\bibfield  {journal}
  {\bibinfo  {journal} {Physics Reports}\ }\textbf {\bibinfo {volume} {209}},\
  \bibinfo {pages} {1} (\bibinfo {year} {1991})}\BibitemShut {NoStop}%
\bibitem [{\citenamefont {Jordan}\ and\ \citenamefont
  {Sukhorukov}(2004)}]{Jordan04}%
  \BibitemOpen
  \bibfield  {author} {\bibinfo {author} {\bibfnamefont {A.~N.}\ \bibnamefont
  {Jordan}}\ and\ \bibinfo {author} {\bibfnamefont {E.~V.}\ \bibnamefont
  {Sukhorukov}},\ }\href {\doibase 10.1103/PhysRevLett.93.260604} {\bibfield
  {journal} {\bibinfo  {journal} {Phys. Rev. Lett.}\ }\textbf {\bibinfo
  {volume} {93}},\ \bibinfo {pages} {260604} (\bibinfo {year}
  {2004})}\BibitemShut {NoStop}%
\bibitem [{\citenamefont {{H\"anggi}}\ \emph {et~al.}(1990)\citenamefont
  {{H\"anggi}}, \citenamefont {{Talkner}},\ and\ \citenamefont
  {{Borkovec}}}]{Hanggi90}%
  \BibitemOpen
  \bibfield  {author} {\bibinfo {author} {\bibfnamefont {P.}~\bibnamefont
  {{H\"anggi}}}, \bibinfo {author} {\bibfnamefont {P.}~\bibnamefont
  {{Talkner}}}, \ and\ \bibinfo {author} {\bibfnamefont {M.}~\bibnamefont
  {{Borkovec}}},\ }\href {\doibase 10.1103/RevModPhys.62.251} {\bibfield
  {journal} {\bibinfo  {journal} {Rev. Mod. Phys.}\ }\textbf {\bibinfo {volume}
  {62}},\ \bibinfo {pages} {251} (\bibinfo {year} {1990})}\BibitemShut
  {NoStop}%
\bibitem [{\citenamefont {{Ben-Jacob}}\ \emph {et~al.}(1982)\citenamefont
  {{Ben-Jacob}}, \citenamefont {{Bergman}}, \citenamefont {{Matkowsky}},\ and\
  \citenamefont {{Schuss}}}]{BenJacob82}%
  \BibitemOpen
  \bibfield  {author} {\bibinfo {author} {\bibfnamefont {E.}~\bibnamefont
  {{Ben-Jacob}}}, \bibinfo {author} {\bibfnamefont {D.~J.}\ \bibnamefont
  {{Bergman}}}, \bibinfo {author} {\bibfnamefont {B.~J.}\ \bibnamefont
  {{Matkowsky}}}, \ and\ \bibinfo {author} {\bibfnamefont {Z.}~\bibnamefont
  {{Schuss}}},\ }\href {\doibase 10.1103/PhysRevA.26.2805} {\bibfield
  {journal} {\bibinfo  {journal} {Phys. Rev. A}\ }\textbf {\bibinfo {volume}
  {26}},\ \bibinfo {pages} {2805} (\bibinfo {year} {1982})}\BibitemShut
  {NoStop}%
\bibitem [{\citenamefont {{Mel'nikov}}(1987)}]{Melnikov87}%
  \BibitemOpen
  \bibfield  {author} {\bibinfo {author} {\bibfnamefont {V.~I.}\ \bibnamefont
  {{Mel'nikov}}},\ }\href@noop {} {\bibfield  {journal} {\bibinfo  {journal}
  {Zh. Eksp. Teor. Fiz.}\ }\textbf {\bibinfo {volume} {93}},\ \bibinfo {pages}
  {2037} (\bibinfo {year} {1987})}\BibitemShut {NoStop}%
\bibitem [{\citenamefont {{Jung}}\ and\ \citenamefont
  {{Risken}}(1984)}]{Jung83}%
  \BibitemOpen
  \bibfield  {author} {\bibinfo {author} {\bibfnamefont {P.}~\bibnamefont
  {{Jung}}}\ and\ \bibinfo {author} {\bibfnamefont {H.}~\bibnamefont
  {{Risken}}},\ }\href {\doibase 10.1007/BF01485833} {\bibfield  {journal}
  {\bibinfo  {journal} {Zeitschrift fur Physik B Condensed Matter}\ }\textbf
  {\bibinfo {volume} {54}},\ \bibinfo {pages} {357} (\bibinfo {year}
  {1984})}\BibitemShut {NoStop}%
\bibitem [{\citenamefont {M{\"a}nnik}\ \emph {et~al.}(2005)\citenamefont
  {M{\"a}nnik}, \citenamefont {Li}, \citenamefont {Qiu}, \citenamefont {Chen},
  \citenamefont {Patel}, \citenamefont {Han},\ and\ \citenamefont
  {Lukens}}]{Lukens:PRB05}%
  \BibitemOpen
  \bibfield  {author} {\bibinfo {author} {\bibfnamefont {J.}~\bibnamefont
  {M{\"a}nnik}}, \bibinfo {author} {\bibfnamefont {S.}~\bibnamefont {Li}},
  \bibinfo {author} {\bibfnamefont {W.}~\bibnamefont {Qiu}}, \bibinfo {author}
  {\bibfnamefont {W.}~\bibnamefont {Chen}}, \bibinfo {author} {\bibfnamefont
  {V.}~\bibnamefont {Patel}}, \bibinfo {author} {\bibfnamefont
  {S.}~\bibnamefont {Han}}, \ and\ \bibinfo {author} {\bibfnamefont {J.~E.}\
  \bibnamefont {Lukens}},\ }\href
  {http://link.aps.org/doi/10.1103/PhysRevB.71.220509} {\bibfield  {journal}
  {\bibinfo  {journal} {Physical Review B}\ }\textbf {\bibinfo {volume} {71}},\
  \bibinfo {pages} {220509} (\bibinfo {year} {2005})}\BibitemShut {NoStop}%
\bibitem [{\citenamefont {Krasnov}\ \emph {et~al.}(2005)\citenamefont
  {Krasnov}, \citenamefont {Bauch}, \citenamefont {Intiso}, \citenamefont
  {H{\"u}rfeld}, \citenamefont {Akazaki}, \citenamefont {Takayanagi},\ and\
  \citenamefont {Delsing}}]{Delsing:PRL05}%
  \BibitemOpen
  \bibfield  {author} {\bibinfo {author} {\bibfnamefont {V.~M.}\ \bibnamefont
  {Krasnov}}, \bibinfo {author} {\bibfnamefont {T.}~\bibnamefont {Bauch}},
  \bibinfo {author} {\bibfnamefont {S.}~\bibnamefont {Intiso}}, \bibinfo
  {author} {\bibfnamefont {E.}~\bibnamefont {H{\"u}rfeld}}, \bibinfo {author}
  {\bibfnamefont {T.}~\bibnamefont {Akazaki}}, \bibinfo {author} {\bibfnamefont
  {H.}~\bibnamefont {Takayanagi}}, \ and\ \bibinfo {author} {\bibfnamefont
  {P.}~\bibnamefont {Delsing}},\ }\href
  {http://link.aps.org/doi/10.1103/PhysRevLett.95.157002} {\bibfield  {journal}
  {\bibinfo  {journal} {Physical Review Letters}\ }\textbf {\bibinfo {volume}
  {95}},\ \bibinfo {pages} {157002} (\bibinfo {year} {2005})}\BibitemShut
  {NoStop}%
\bibitem [{\citenamefont {Kivioja}\ \emph {et~al.}(2005)\citenamefont
  {Kivioja}, \citenamefont {Nieminen}, \citenamefont {Claudon}, \citenamefont
  {Buisson}, \citenamefont {Hekking},\ and\ \citenamefont
  {Pekola}}]{Pekola:PRL05}%
  \BibitemOpen
  \bibfield  {author} {\bibinfo {author} {\bibfnamefont {J.~M.}\ \bibnamefont
  {Kivioja}}, \bibinfo {author} {\bibfnamefont {T.~E.}\ \bibnamefont
  {Nieminen}}, \bibinfo {author} {\bibfnamefont {J.}~\bibnamefont {Claudon}},
  \bibinfo {author} {\bibfnamefont {O.}~\bibnamefont {Buisson}}, \bibinfo
  {author} {\bibfnamefont {F.~W.~J.}\ \bibnamefont {Hekking}}, \ and\ \bibinfo
  {author} {\bibfnamefont {J.~P.}\ \bibnamefont {Pekola}},\ }\href
  {http://link.aps.org/doi/10.1103/PhysRevLett.94.247002} {\bibfield  {journal}
  {\bibinfo  {journal} {Physical Review Letters}\ }\textbf {\bibinfo {volume}
  {94}},\ \bibinfo {pages} {247002} (\bibinfo {year} {2005})}\BibitemShut
  {NoStop}%
\bibitem [{\citenamefont {Krasnov}\ \emph {et~al.}(2007)\citenamefont
  {Krasnov}, \citenamefont {Golod}, \citenamefont {Bauch},\ and\ \citenamefont
  {Delsing}}]{Delsing:PRB07}%
  \BibitemOpen
  \bibfield  {author} {\bibinfo {author} {\bibfnamefont {V.~M.}\ \bibnamefont
  {Krasnov}}, \bibinfo {author} {\bibfnamefont {T.}~\bibnamefont {Golod}},
  \bibinfo {author} {\bibfnamefont {T.}~\bibnamefont {Bauch}}, \ and\ \bibinfo
  {author} {\bibfnamefont {P.}~\bibnamefont {Delsing}},\ }\href
  {http://link.aps.org/doi/10.1103/PhysRevB.76.224517} {\bibfield  {journal}
  {\bibinfo  {journal} {Physical Review B}\ }\textbf {\bibinfo {volume} {76}},\
  \bibinfo {pages} {224517} (\bibinfo {year} {2007})}\BibitemShut {NoStop}%
\bibitem [{\citenamefont {Murphy}\ \emph {et~al.}(2013)\citenamefont {Murphy},
  \citenamefont {Weinberg}, \citenamefont {Aref}, \citenamefont {Coskun},
  \citenamefont {Vakaryuk}, \citenamefont {Levchenko},\ and\ \citenamefont
  {Bezryadin}}]{Levchenko:PRL13}%
  \BibitemOpen
  \bibfield  {author} {\bibinfo {author} {\bibfnamefont {A.}~\bibnamefont
  {Murphy}}, \bibinfo {author} {\bibfnamefont {P.}~\bibnamefont {Weinberg}},
  \bibinfo {author} {\bibfnamefont {T.}~\bibnamefont {Aref}}, \bibinfo {author}
  {\bibfnamefont {U.~C.}\ \bibnamefont {Coskun}}, \bibinfo {author}
  {\bibfnamefont {V.}~\bibnamefont {Vakaryuk}}, \bibinfo {author}
  {\bibfnamefont {A.}~\bibnamefont {Levchenko}}, \ and\ \bibinfo {author}
  {\bibfnamefont {A.}~\bibnamefont {Bezryadin}},\ }\href
  {http://link.aps.org/doi/10.1103/PhysRevLett.110.247001} {\bibfield
  {journal} {\bibinfo  {journal} {Physical Review Letters}\ }\textbf {\bibinfo
  {volume} {110}},\ \bibinfo {pages} {247001} (\bibinfo {year}
  {2013})}\BibitemShut {NoStop}%
\bibitem [{\citenamefont {Challis}\ and\ \citenamefont
  {Jack}(2013)}]{Challis:PRE13}%
  \BibitemOpen
  \bibfield  {author} {\bibinfo {author} {\bibfnamefont {K.~J.}\ \bibnamefont
  {Challis}}\ and\ \bibinfo {author} {\bibfnamefont {M.~W.}\ \bibnamefont
  {Jack}},\ }\href {http://link.aps.org/doi/10.1103/PhysRevE.87.052102}
  {\bibfield  {journal} {\bibinfo  {journal} {Physical Review E}\ }\textbf
  {\bibinfo {volume} {87}},\ \bibinfo {pages} {052102} (\bibinfo {year}
  {2013})}\BibitemShut {NoStop}%
\bibitem [{\citenamefont {{Kramers}}(1940)}]{Kramers40}%
  \BibitemOpen
  \bibfield  {author} {\bibinfo {author} {\bibfnamefont {H.~A.}\ \bibnamefont
  {{Kramers}}},\ }\href {\doibase 10.1016/S0031-8914(40)90098-2} {\bibfield
  {journal} {\bibinfo  {journal} {Physica}\ }\textbf {\bibinfo {volume} {7}},\
  \bibinfo {pages} {284} (\bibinfo {year} {1940})}\BibitemShut {NoStop}%
\end{thebibliography}

\begin{thebibliography}{3}
\expandafter\ifx\csname natexlab\endcsname\relax\def\natexlab#1{#1}\fi
\expandafter\ifx\csname bibnamefont\endcsname\relax
  \def\bibnamefont#1{#1}\fi
\expandafter\ifx\csname bibfnamefont\endcsname\relax
  \def\bibfnamefont#1{#1}\fi
\expandafter\ifx\csname citenamefont\endcsname\relax
  \def\citenamefont#1{#1}\fi
\expandafter\ifx\csname url\endcsname\relax
  \def\url#1{\texttt{#1}}\fi
\expandafter\ifx\csname urlprefix\endcsname\relax\def\urlprefix{URL }\fi
\providecommand{\bibinfo}[2]{#2}
\providecommand{\eprint}[2][]{\url{#2}}

\bibitem[{\citenamefont{{Risken}}(1989)}]{RiskenS}
\bibinfo{author}{\bibfnamefont{H.}~\bibnamefont{{Risken}}},
  \emph{\bibinfo{title}{The Fokker-Planck Equation, 2nd ed.}}
  (\bibinfo{publisher}{Springer-Verlag, Berlin, Heidelberg},
  \bibinfo{year}{1989}).

\bibitem[{\citenamefont{Flindt et~al.}(2004)\citenamefont{Flindt, Novotn\'{y},
  and Jauho}}]{Flindt2004S}
\bibinfo{author}{\bibfnamefont{C.}~\bibnamefont{Flindt}},
  \bibinfo{author}{\bibfnamefont{T.}~\bibnamefont{Novotn\'{y}}},
  \bibnamefont{and} \bibinfo{author}{\bibfnamefont{A.-P.} \bibnamefont{Jauho}},
  \bibinfo{journal}{Phys. Rev. B} \textbf{\bibinfo{volume}{70}},
  \bibinfo{pages}{205334} (\bibinfo{year}{2004}).

\bibitem[{\citenamefont{{Golubev} et~al.}(2010)\citenamefont{{Golubev},
  {Marthaler}, {Utsumi}, and {Sch\"on}}}]{Golubev10S}
\bibinfo{author}{\bibfnamefont{D.~S.} \bibnamefont{{Golubev}}},
  \bibinfo{author}{\bibfnamefont{M.}~\bibnamefont{{Marthaler}}},
  \bibinfo{author}{\bibfnamefont{Y.}~\bibnamefont{{Utsumi}}}, \bibnamefont{and}
  \bibinfo{author}{\bibfnamefont{G.}~\bibnamefont{{Sch\"on}}},
  \bibinfo{journal}{Phys. Rev. B} \textbf{\bibinfo{volume}{81}},
  \bibinfo{pages}{184516} (\bibinfo{year}{2010}).

\end{thebibliography}

%

\newpage
\newpage
\setcounter{page}{1}
\setcounter{equation}{0}

\onecolumngrid

\title{Supplemental Material for ``Voltage noise, switching rates, and
multiple phase-slips in moderately damped Josephson junctions''}

\author{Martin \v{Z}onda}


\author{Wolfgang Belzig}


\author{Tom\'{a}\v{s} Novotn\'{y} }

%

\pacs{}
\maketitle
\onecolumngrid

Aim of this supplemental material is to provide an introduction to
a reader not familiar with the methods used in our paper, namely the
Matrix Continued-Fraction (MCF) method and the full counting statistics
(FCS).

\subsection*{Matrix Continued-Fraction Method}

In the paper we discussed a numerical solution of the dimensionless
Langevin equations 
\begin{equation}
\begin{split}\frac{\partial v(\tau)}{\partial\tau} & =i_{b}-\gamma v(\tau)-\sin\phi(\tau)+\zeta(\tau),\\
v(t) & =\partial\phi/\partial\tau,
\end{split}
\label{eq:Langevin-dimensionless}
\end{equation}
with the associated Fokker-Planck equation for the probability distribution
function $W(\phi,v,\tau)$ 
\begin{equation}
\begin{split}\dfrac{\partial}{\partial\tau}W(\phi,v;\tau) & =-v\dfrac{\partial}{\partial\phi}W+\dfrac{\partial}{\partial v}\left(\gamma v+\sin\phi-i_{b}+\gamma\Theta\dfrac{\partial}{\partial v}\right)W\\
 & \equiv L_{\mathrm{FP}}W(\phi,v;\tau).
\end{split}
\label{eq:Fokker-Planck}
\end{equation}
Our first goal was to calculate the mean voltage 
\begin{equation}
\langle v\rangle=\int\limits _{0}^{2\pi}\mathrm{d}\phi\int\limits _{-\infty}^{\infty}\mathrm{d}vvW_{\mathrm{stat}}(\phi,v),
\end{equation}
and the (zero-frequency) voltage noise
\begin{equation}
S=\int\limits _{-\infty}^{\infty}\mathrm{d}\tau\big(\langle v(\tau)v(0)\rangle-\langle v(\tau)\rangle\langle v(0)\rangle\big)\label{eq:Noise}
\end{equation}
in the stationary state $W_{\mathrm{stat}}(\phi,v)\equiv\lim_{\tau\to\infty}W(\phi,v;\tau)$
determined by the $2\pi$-periodic solution (in $\phi$) of the equation
\begin{equation}
L_{\mathrm{FP}}W_{\mathrm{stat}}(\phi,v)=0.\label{eq:LWStat}
\end{equation}
We have used the MCF method \cite{RiskenS} to obtain the numerical
solution for the stationary distribution function and our explanation
closely follows that work. The operator $L_{\mathrm{FP}}$ can be
partitioned into irreversible $L_{i}$ and reversible $L_{r}$ operators
\begin{eqnarray}
L_{i} & = & \gamma\dfrac{\partial}{\partial v}\left(v+\Theta\dfrac{\partial}{\partial v}\right),\\
L_{r} & = & -v\dfrac{\partial}{\partial\phi}W-U'(\phi)\dfrac{\partial}{\partial v},
\end{eqnarray}
with 
\begin{equation}
U(\phi)=i_{b}\phi+\cos\phi.\label{eq:Potential}
\end{equation}
The irreversible operator $L_{i}$ can be mapped onto the Hamilton
operator of the linear harmonic oscillator via a suitable similarity
transformation
\begin{equation}
\widetilde{L_{i}}=\exp\left[\frac{v^{2}}{4\Theta}\right]L_{i}\exp\left[-\frac{v^{2}}{4\Theta}\right]=-\gamma b^{\dagger}b,\label{eq:Li}
\end{equation}
where the creation $b^{\dagger}$ and annihilation $b$ operators
were introduced 
\begin{eqnarray}
b^{\dagger} & = & -\sqrt{\Theta}\dfrac{\partial}{\partial v}+\frac{v}{2\sqrt{\Theta}},\\
b\, & = & \;\sqrt{\Theta}\dfrac{\partial}{\partial v}+\frac{v}{2\sqrt{\Theta}}.
\end{eqnarray}
For the reversible operator we consequently get
\begin{eqnarray}
\widetilde{L_{r}} & = & \exp\left[\frac{v^{2}}{4\Theta}\right]L_{r}\exp\left[-\frac{v^{2}}{4\Theta}\right]=-b^{\dagger}D_{2}-bD_{1},
\end{eqnarray}
with $\phi$- dependent operators
\begin{eqnarray}
D_{1} & = & \sqrt{\Theta}\dfrac{\partial}{\partial\phi},\\
D_{2} & = & \:\sqrt{\Theta}\dfrac{\partial}{\partial\phi}-\frac{U'(\phi)}{\sqrt{\Theta}}.
\end{eqnarray}
Altogether we have 
\begin{equation}
L_{\mathrm{FP}}=\exp\left[-\frac{v^{2}}{4\Theta}\right]\left(-\gamma b^{\dagger}b-b^{\dagger}D_{2}-bD_{1}\right)\exp\left[\frac{v^{2}}{4\Theta}\right].\label{eq:LFP}
\end{equation}
After these manipulations it is convenient to expand the $v$ part
of distribution function $W(\phi,v;\tau)$ into the Hermite oscillator
functions $\psi_{n}(v)$ 
\begin{equation}
W(\phi,v;\tau)=\psi_{0}(v)\sum_{n}c_{n}(\phi;\tau)\psi_{n}(v),\label{eq:W}
\end{equation}
which are given by
\begin{eqnarray*}
\psi_{0}(v) & = & e^{-\frac{1}{2}\left(\kappa v\right)^{2}}/\sqrt{\kappa\sqrt{\pi}},\\
\psi_{n}(v) & = & \left(b^{\dagger}\right)^{n}\psi_{0}(v)/\sqrt{n!},
\end{eqnarray*}
with $\kappa=1/\sqrt{2\Theta}$ or in terms of Hermite polynomials
$H_{n}\left(\kappa v\right)$ used in quantum mechanics 
\begin{equation}
\psi_{n}(v)=H_{n}\left(\kappa v\right)e^{-\frac{1}{2}\left(\kappa v\right)^{2}}/\sqrt{n!2^{n}\kappa\sqrt{\pi}}.
\end{equation}
The oscillator functions are the eigenfunctions of the irreversible
part of the $L_{\mathrm{FP}}$ operator and have the correct boundary
conditions ($\lim\limits _{v\rightarrow\pm\infty}W(\phi,v;\tau)=0$).
Consequently,
\begin{eqnarray}
\int\limits _{-\infty}^{\infty}dv\, W(\phi,v;\tau) & = & c_{0}(\phi;\tau),\\
\int\limits _{-\infty}^{\infty}dv\, vW(\phi,v;\tau) & = & \sqrt{\Theta}c_{1}(\phi;\tau),\label{eq:mean-voltage}
\end{eqnarray}
and the initial values of the $\phi$-part coefficients read
\begin{equation}
c_{n}(\phi;0)=\int\limits _{-\infty}^{\infty}dv\,\psi_{n}\psi_{0}^{-1}W(\phi,v;0).\label{eq:InVal}
\end{equation}
Using Eq.~\eqref{eq:Li}-\eqref{eq:W} the Brinkman hierarchy, equivalent
to the Fokker-Planck equation \eqref{eq:Fokker-Planck}, can be constructed
\begin{equation}
-\sqrt{m}D_{2}c_{m-1}(\phi;\tau)-\gamma mc_{m}(\phi;\tau)-\sqrt{m+1}D_{1}c_{m+1}(\phi,;\tau)=\dfrac{\partial}{\partial\tau}c_{m}(\phi;\tau).\label{eq:BrinkmanH}
\end{equation}

\subsubsection*{Stationary Distribution Function and Average Voltage }

Because of the chosen potential Eq.~\eqref{eq:Potential} the Fokker-Planck
operator Eq.~\eqref{eq:LFP} commutes with the translation operator
$T$ defined by
\begin{equation}
T\, W(\phi,v;\tau)=W(\phi+2\pi,v;\tau),
\end{equation}
therefore the eigenfunctions $\varphi_{n}(k,\phi,v)$ of operator
$L_{\mathrm{FP}}$ and its adjoint operator $L_{\mathrm{FP}}^{+}$
\begin{eqnarray*}
L_{\mathrm{FP}}\varphi_{n}(k,\phi,v) & = & \lambda_{n}(k)\varphi_{n}(k,\phi,v),\\
L_{\mathrm{FP}}^{+}\varphi_{n}^{+}(k,\phi,v) & = & \lambda_{n}(k)\varphi_{n}^{+}(k,\phi,v),
\end{eqnarray*}
with $k\in(-1/2,1/2]$ restricted to the first Brillouin zone can
be written in the form of the Bloch waves
\begin{eqnarray}
\varphi_{n}(k,\phi,v) & = & e^{-ik\phi}u_{n}\left(k,\phi,v\right),\; u_{n}(k,\phi,v)=u_{n}(k,\phi+2\pi,v),\nonumber \\
\varphi_{n}^{+}(k,\phi,v) & = & e^{ik\phi}u_{n}^{+}\left(k,\phi,v\right),\; u_{n}^{+}(k,\phi,v)=u_{n}^{+}(k,\phi+2\pi,v).\label{eq:LambdaExp}
\end{eqnarray}
 Consequently, the solutions of the Eq.~\eqref{eq:LWStat} can be
chosen to be
\begin{equation}
W_{\mathrm{stat}}(\phi,v)=e^{-ik\phi}u\left(k,\phi,v\right),
\end{equation}
with $e^{-i2\pi k},\,-1/2<k\leq1/2$, being the eigenvalues of the
translation operator $T$. The stationary expansion coefficients $c_{n}(\phi)\equiv\lim_{\tau\rightarrow\infty}c_{n}(\phi;\tau)$
must therefore have the form
\begin{equation}
c_{n}(\phi)=e^{-ik\phi}u_{n}(k,\phi),\quad u_{n}(k,\phi)=u_{n}(k,\phi+2\pi).
\end{equation}
Note that the stationary form of the Brinkman hierarchy Eq.~\eqref{eq:BrinkmanH}
reads
\begin{eqnarray}
\sqrt{1}D_{1}c_{1}(\phi) & = & 0\nonumber \\
\sqrt{1}D_{2}c_{0}(\phi)+1\gamma c_{1}(\phi)+\sqrt{2}D_{1}c_{2}(\phi) & = & 0\label{eq:BrinkmanStat}\\
\sqrt{2}D_{2}c_{1}(\phi)+2\gamma c_{2}(\phi)+\sqrt{3}D_{1}c_{3}(\phi) & = & 0\nonumber \\
 & \vdots\nonumber 
\end{eqnarray}
from which it is obvious that $c_{1}(\phi)=c_{1}=\mathrm{const.}$
Since $c_{1}$ is related by Eq.~\eqref{eq:mean-voltage} to the
mean voltage being generically non-zero, the constancy of $c_{1}$
implies $k=0$ for the stationary solution. Thus, the stationary distribution
function is periodic
\begin{equation}
W_{\mathrm{stat}}(\phi,v)=W_{\mathrm{stat}}(\phi+2\phi,v),\quad c_{n}(\phi)=c_{n}(\phi+2\pi)
\end{equation}
and can be normalized in one period

\begin{eqnarray}
\int\limits _{-\infty}^{\infty}dv\,\intop_{0}^{2\pi}d\phi\, W_{\mathrm{stat}}(\phi,v) & = & \intop_{0}^{2\pi}d\phi c_{0}(\phi)=1,\label{eq:Normal}\\
\left\langle v\right\rangle =\int\limits _{-\infty}^{\infty}dv\,\intop_{0}^{2\pi}d\phi\, vW_{\mathrm{stat}}(\phi,v) & = & \sqrt{\Theta}\intop_{0}^{2\pi}d\phi c_{1}(\phi)=\sqrt{\Theta}2\pi c_{1}.
\end{eqnarray}
To solve Eq.~\eqref{eq:LWStat} for the periodic coefficients $c_{m}(\phi)$
we have used the Fourier expansion 
\begin{equation}
c_{m}(\phi)=\frac{1}{\sqrt{2\pi}}\sum_{p}c_{m}^{p}\, e^{ip\phi},\label{eq:ExpFour}
\end{equation}
allowing us to define the matrix elements of operators $\mathcal{D_{\mathit{m}}^{\mathit{+}}},\mathcal{D_{\mathit{m}}},\mathcal{D_{\mathit{m}}^{\mathit{-}}}$
\begin{eqnarray}
\mathcal{\left(D_{\mathit{m}}^{\mathit{+}}\right)}^{pq} & \equiv & -\frac{\sqrt{m+1}}{2\pi}\,\int\limits _{0}^{2\pi}d\phi\, e^{-ip\phi}D_{1}e^{iq\phi}=-i\sqrt{m+1}\sqrt{\Theta}p\,\delta_{p,q},\nonumber \\
\mathcal{\left(D_{\mathit{m}}\right)}^{pq} & \equiv & -\gamma m\,\delta_{p,q},\\
\mathcal{\left(D_{\mathit{m}}^{\mathit{-}}\right)}^{pq} & \equiv & -\frac{\sqrt{m}}{2\pi}\,\int\limits _{0}^{2\pi}d\phi\, e^{-ip\phi}D_{2}e^{iq\phi}=-i\sqrt{m}\sqrt{\Theta}\left[\left(p+i\frac{i_{b}}{\Theta}\right)\,\delta_{p,q}+\frac{\delta_{p,q-1}-\delta_{p,q+1}}{2\Theta}\right],\nonumber 
\end{eqnarray}
which can be used to recast the Eq.~\eqref{eq:BrinkmanStat} into
the form of a vector tridiagonal recurrence relation
\begin{equation}
\mathcal{D_{\mathit{m}}^{\mathit{-}}}\mathbf{c}_{m-1}+\mathcal{D_{\mathit{m}}}\mathbf{c}_{m}+\mathcal{D_{\mathit{m}}^{\mathit{+}}}\mathbf{c}_{m+1}=0,\label{eq:ReRe}
\end{equation}
where $\mathbf{c}_{m}$ is a time-independent vector of expansion
coefficients $c_{m}^{p}$ from Eq.~\eqref{eq:ExpFour}
\begin{equation}
\mathbf{c}_{m}=\left(\begin{array}{c}
\vdots\\
c_{m}^{-1}\\
c_{m}^{0}\\
c_{m}^{1}\\
\vdots
\end{array}\right).
\end{equation}
To solve the relation \eqref{eq:ReRe} we defined matrices $\mathcal{S}_{m}$
obeying 
\begin{equation}
\mathbf{c}_{m+1}=\mathcal{S}_{m}\mathbf{c}_{m},\quad\mathbf{c}_{m}=\mathcal{S}_{m}^{-1}\mathbf{c}_{m+1},\label{eq:S}
\end{equation}
which transforms the Eq.~\eqref{eq:ReRe} into 
\begin{equation}
\mathcal{D_{\mathit{m}}^{\mathit{-}}}\mathbf{\mathcal{S}_{\mathrm{\mathit{m-\mathrm{1}}}}^{-\mathrm{1}}c}_{m}+\mathcal{D_{\mathit{m}}}\mathbf{c}_{m}+\mathcal{D_{\mathit{m}}^{\mathit{+}}}\mathcal{S}_{m}\mathbf{c}_{m}=0,\label{eq:ReRe-1}
\end{equation}
and consequently a matrix continued-fraction structure can be constructed
\begin{equation}
\mathcal{S}_{m-1}=-\left(\mathcal{D_{\mathit{m}}}+\mathcal{D_{\mathit{m}}^{\mathit{+}}\mathcal{S}_{\mathit{m}}}\right)^{-1}\mathcal{D_{\mathit{m}}^{\mathit{-}}}.
\end{equation}
By truncating the recurrence at $m=M$, i.e., setting $\mathbf{c}_{m>M}=0$
in Eq.~\eqref{eq:ReRe} and using the normalization condition \eqref{eq:Normal}
together with the fact that the coefficient $c_{1}$ is constant we
obtain for the vector $\mathbf{c}_{0}$
\begin{equation}
c_{0}^{p}=\frac{1}{\sqrt{2\pi}}\frac{\left(\mathcal{S}_{0}^{-1}\right)^{p0}}{\left(\mathcal{S}_{0}^{-1}\right)^{00}}.
\end{equation}
All other vectors $\mathbf{c}_{m}$ follow from Eq.~\eqref{eq:S}.
In particular, the average voltage reads
\begin{equation}
\left\langle v\right\rangle =\sqrt{\Theta}2\pi c_{1}\equiv\sqrt{2\pi\Theta}c_{1}^{0}=\frac{\sqrt{\Theta}}{\left(\mathcal{S}_{0}^{-1}\right)^{00}}.
\end{equation}

\subsubsection*{Voltage noise}

Computation of the voltage noise is more complicated. The general
formula for the frequency dependent voltage noise reads 
\begin{equation}
S(\omega)=\int\limits _{-\infty}^{\infty}\mathrm{d}\tau e^{i\omega\tau}\big(\langle v(\tau)v(0)\rangle-\langle v(\tau)\rangle\langle v(0)\rangle\big)\label{eq:NoiseOmega}
\end{equation}
and the voltage autocorrelation function can be expressed as \cite[Sec.~7.2]{RiskenS}
\begin{eqnarray}
\langle v(\tau)v(0)\rangle & = & \int\limits _{0}^{2\pi}\mathrm{d}\phi\int\limits _{-\infty}^{\infty}\mathrm{d}v\, ve^{|\tau|L_{FP}}vW_{\mathrm{stat}}(\phi,v).
\end{eqnarray}
After introducing the convergence factors $\omega\rightarrow\omega+i0$
for $\tau>0$ and $\omega\rightarrow\omega-i0$ for $\tau<0$ we get
\begin{equation}
S(\omega)=\int\limits _{0}^{2\pi}\mathrm{d}\phi\int\limits _{-\infty}^{\infty}\mathrm{d}v\, v\left(\frac{1}{i\omega-L_{\mathrm{FP}}}-\frac{1}{i\omega+L_{\mathrm{FP}}}\right)vW_{\mathrm{stat}}(\phi,v).
\end{equation}
Since we are interested in the limit $\omega\rightarrow0$ and $L_{\mathrm{FP}}$
is singular (due to the existence of the stationary state), performing
the limit is somewhat tricky. It can be done, however, as explained
in detail in Ref.~\cite[Sec.~IIIB]{Flindt2004S} and the the voltage
noise can be evaluated as 

\begin{equation}
S=-2\int\limits _{0}^{2\pi}\mathrm{d}\phi\int\limits _{-\infty}^{\infty}\mathrm{d}vvR(\phi,v)
\end{equation}
with the help of an auxiliary quantity $R(\phi,v)$ (pseudoinverse
of the Fokker-Planck operator) satisfying the equation 
\begin{eqnarray}
L_{\mathrm{FP}}R(\phi,v) & = & (v-\left\langle v\right\rangle )W_{\mathrm{stat}}(\phi,v),\label{eq:aux}
\end{eqnarray}
and conditions $R(\phi+2\pi,v)=R(\phi,v)$ (periodicity) and $\int\limits _{0}^{2\pi}\mathrm{d}\phi\int\limits _{-\infty}^{\infty}\mathrm{d}vR(\phi,v)=0$
(fixing one out of infinitely many solutions of Eq.~\eqref{eq:aux},
see \cite[Sec.~IIIE]{Flindt2004S}). We have obtained the numerical
solution of Eq.~\eqref{eq:aux} analogously to the solution of Eq.~\eqref{eq:LWStat}
--- the main difference is that for Eq.~\eqref{eq:aux} the vector
tridiagonal recurrence relation has a right-hand side
\begin{equation}
\mathcal{D_{\mathit{m}}^{\mathit{-}}}\mathbf{a}_{m-1}+\mathcal{D_{\mathit{m}}}\mathbf{a}_{m}+\mathcal{D_{\mathit{m}}^{\mathit{+}}}\mathbf{a}_{m+1}=\alpha_{m},\label{eq:ReReRS}
\end{equation}
with $\mathbf{a}_{m}$ being a time-independent vector of expansion
coefficients $a_{m}^{p}$ of $R(\phi,v)$ obtained in the same procedure
as coefficients $c_{m}^{p}$ in Eq.~\eqref{eq:ExpFour} for $W_{\mathrm{stat}}(\phi,v)$
and
\begin{equation}
\alpha_{m}=\sqrt{\Theta m}\mathbf{c}_{m-1}-\left\langle v\right\rangle \mathbf{c}_{m}+\sqrt{\Theta(m+1)}\mathbf{c}_{m+1}.
\end{equation}
Introducing the correction vectors $g_{m}$ satisfying 
\begin{equation}
\mathbf{a}_{m+1}=\mathcal{S}_{m}\mathbf{a}_{m}+g_{m},\label{eq:Sa}
\end{equation}
the Eq.~\eqref{eq:ReReRS} gives a recurrent prescription for their
evaluation 
\begin{equation}
g_{m-1}=-\left(\mathcal{D_{\mathit{m}}}+\mathcal{D_{\mathit{m}}^{\mathit{+}}\mathcal{S}_{\mathit{m}}}\right)^{-1}\left(\mathcal{D_{\mathit{m}}^{\mathit{+}}}g_{m}-\alpha_{m}\right).
\end{equation}
After truncation of Eq.~\eqref{eq:ReReRS} at $m=M$ and using the
proper normalization conditions together with Eq.~\eqref{eq:Sa}
this recurrent relation is used to obtain the auxiliary quantity $R(\phi,v)$
and, consequently, the voltage noise.

\subsubsection*{Non-stationary solution}

As we are interested in the statistics of $2\pi n$ phase-slips we
have to consider also the non-periodic solutions of the Fokker-Planck
equation in the non-stationary case. Recalling the expansion Eq.~\eqref{eq:W}
and the Brinkman hierarchy Eq.~\eqref{eq:BrinkmanH} it is clear
that to make use of the MCF method we need a complete set of functions
$\varphi^{p}(\phi)$ in which the coefficients $c_{m}(\phi;\tau)$
can be expanded. One of the possibilities is to use the eigenfunctions
Eq.~\eqref{eq:LambdaExp} or equivalently make use of the Floquet
theorem as an ansatz for the non-periodic solution
\begin{equation}
W(\phi,v;\tau)=\intop_{-\frac{1}{2}}^{\frac{1}{2}}dk\,\mathcal{W}(k,\phi,v;\tau)e^{-ik\phi},
\end{equation}
where $\mathcal{W}(k,\phi,v;\tau)$ is periodic in $\phi$ with period
$2\pi$ and $k$ is restricted to the first Brillouin zone. Similarly
as was done before for the distribution function the function $\mathcal{W}(k,\phi,v;\tau)$
can be expanded in 
\begin{equation}
\mathcal{W}(k,\phi,v;\tau)=\psi_{0}(v)\sum_{n}\sum_{p}c_{n}^{p}(k;\tau)e^{ip\phi}\psi_{n}(v),
\end{equation}
yielding the complete set of functions in which the coefficients are
expanded 
\begin{equation}
\varphi^{p}(k,\phi)=\frac{1}{\sqrt{2\pi}}e^{i(p-k)\phi}.\label{eq:ComSet}
\end{equation}
Using time-dependent vectors $\mathbf{c}_{m}(k;\tau)$ of expansion
coefficients $c_{m}^{p}(k,\tau)$, Eq.~\eqref{eq:BrinkmanH} changes
to
\begin{equation}
\mathcal{D_{\mathit{m}}^{\mathit{-}}}\mathbf{c}_{m-1}(k;\tau)+\mathcal{D_{\mathit{m}}}\mathbf{c}_{m}(k;\tau)+\mathcal{D_{\mathit{m}}^{\mathit{+}}}\mathbf{c}_{m+1}(k;\tau)=\mathbf{\dot{c}}_{m}(k;\tau),\label{eq:ReReTau}
\end{equation}
with 
\begin{align}
\mathcal{\left(D_{\mathit{m}}^{\mathit{+}}\right)}^{pq} & =-\frac{\sqrt{m+1}}{2\pi}\,\int\limits _{0}^{2\pi}d\phi\, e^{-i(p-k)\phi}D_{1}e^{i(p-k)\phi}=-i\sqrt{m+1}\sqrt{\Theta}(p-k)\,\delta_{p,q},\\
\mathcal{\left(D_{\mathit{m}}\right)}^{pq} & =-\gamma m\,\delta_{p,q},\\
\mathcal{\left(D_{\mathit{m}}^{\mathit{-}}\right)}^{pq} & =-\frac{\sqrt{m}}{2\pi}\,\int\limits _{0}^{2\pi}d\phi\, e^{-i(p-k)\phi}D_{2}e^{i(p-k)\phi}=-i\sqrt{m}\sqrt{\Theta}\left[\left(p-k+i\frac{i_{b}}{\Theta}\right)\,\delta_{p,q}+\frac{\delta_{p,q-1}-\delta_{p,q+1}}{2\Theta}\right].
\end{align}
One way to solve this initial value problem is to use the Laplace
transform 
\begin{equation}
\mathbf{\tilde{c}}_{m}(k;s)=\int\limits _{0}^{\infty}d\tau\,\mathbf{c}_{m}(k;\tau)\, e^{-s\tau},\label{eq:Laplace}
\end{equation}
 which turns Eq.~\eqref{eq:BrinkmanH} into
\begin{equation}
\mathcal{D_{\mathit{m}}^{\mathit{-}}}\tilde{\mathbf{c}}_{m-1}(k;s)+\mathcal{\tilde{D}_{\mathit{m}}}\tilde{\mathbf{c}}_{m}(k;s)+\mathcal{D_{\mathit{m}}^{\mathit{+}}}\tilde{\mathbf{c}}_{m+1}(k;s)=-\mathbf{c}_{m}(k;0),\label{eq:BrinkmanD}
\end{equation}
with 
\begin{equation}
\mathcal{\tilde{D}_{\mathit{m}}}=\mathcal{D_{\mathit{m}}}-sI.
\end{equation}
This equation can be solved analogously to the solution of Eq.~\eqref{eq:ReReRS}
and the resulting $s$-dependent quantities can be inverse Laplace
transformed to the time domain. Alternatively, one can use the homogeneous
version of Eq.~\eqref{eq:BrinkmanD} for determining the eigenvalues
of the Fokker-Planck operator from the condition 
\begin{equation}
\mathrm{Det}\left[\mathcal{D_{\mathit{m}}}-\lambda I+\mathcal{D_{\mathit{m}}^{\mathit{-}}S_{\mathit{m-1}}^{\mathit{-1}}}(\lambda)+\mathcal{D_{\mathit{m}}^{\mathit{+}}S_{\mathit{m}}}(\lambda)\right]=0
\end{equation}
 and, consequently, for finding the nonstationary solution by the
spectral decomposition 
\begin{equation}
W(\phi,v;\tau)=\intop_{-1/2}^{1/2}dk\sum_{n}e^{-ik\phi}u_{n}\left(k,\phi,v\right)e^{\lambda_{n}(k)\tau}.
\end{equation}
The advantage of the eigenfunction expansion is that the transition
probability from state $\phi',v'$ to $\phi,v$ has a simple form
\cite[Sec.~11.7]{RiskenS} 
\begin{equation}
P(\phi,v;\tau|\phi',v',0)=\int\limits _{-1/2}^{1/2}dk\sum_{n}u_{n}^{+}(k,\phi',v')u_{n}(k,\phi,v)e^{-ik(\phi-\phi')}e^{\lambda_{n}(k)\tau},
\end{equation}
whose long time $\tau\to\infty$ asymptotics is easily determined
as (we assume that the eigenvalue $\lambda_{0}(k)$ with the highest
real part corresponding to the stationary solution with $\lambda_{0}(0)=0$
is separated by a finite gap from other eigenvalues $\Re[\lambda_{n}(k)-\lambda_{0}(k)]<0$
for $n\geq1$) 
\begin{equation}
P(\phi,v;\tau\to\infty|\phi',v',0)\approx\int\limits _{-1/2}^{1/2}dk\, u_{0}^{+}(k,\phi',v')u_{0}(k,\phi,v)e^{-ik(\phi-\phi')}e^{\lambda_{0}(k)\tau},
\end{equation}
or
\begin{equation}
W(\phi,v;\tau\to\infty)=\int\limits _{-1/2}^{1/2}dk\, I_{0}(k)u_{0}(k,\phi,v)e^{-ik\phi}e^{\lambda_{0}(k)\tau},\label{eq:L0exp}
\end{equation}
with $I_{0}(k)$ being determined solely by the initial condition.

\subsection*{Full Counting Statistics}

As pointed out in the paper an alternative method for the evaluation
of the voltage cumulants is to use the full counting statistics (FCS)
approach pioneered in this context in Ref.~\cite{Golubev10S}. The
aim of that method is the calculation of the $k$-dependent ($k$
is the \emph{counting-field}) cumulant generating function (CGF) $F(k;\tau)\equiv\ln\int\limits _{-\infty}^{\infty}\mathrm{d}\phi e^{ik\phi}\intop_{-\infty}^{\infty}dvW(\phi,v;\tau)$
from a non-stationary solution $W(\phi,v;\tau)$ of Eq.~\eqref{eq:Fokker-Planck}
\cite[Sec.~11.7]{RiskenS}. Using the Floquet theorem and Eq.~\eqref{eq:L0exp}
once again the CGF can be written as

\begin{eqnarray}
\exp[F(k;\tau\to\infty)] & = & \int_{-\infty}^{\infty}d\phi\intop_{-\infty}^{\infty}dvW(\phi,v;\tau\to\infty)e^{ik\phi}\nonumber \\
 & \approx & \int_{-\infty}^{\infty}d\phi\intop_{-\infty}^{\infty}dv\intop_{-\frac{1}{2}}^{\frac{1}{2}}dl\, I_{0}(l)u_{0}(l,\phi,v)e^{\lambda_{0}(l)\tau}e^{i(k-l)\phi}\nonumber \\
 & = & \frac{1}{\sqrt{2\pi}}\int\limits _{-\infty}^{\infty}d\phi\intop_{-\frac{1}{2}}^{\frac{1}{2}}dl\, I_{0}(l)\sum_{p}c_{0}^{p}(l)e^{i(p-l+k)\phi}e^{\lambda_{0}(l)\tau}\nonumber \\
 & \text{=} & \frac{1}{\sqrt{2\pi}}\intop_{-\frac{1}{2}}^{\frac{1}{2}}dl\, I_{0}(l)e^{\lambda_{0}(l)\tau}\sum_{p}c_{0}^{p}(l)\lim_{N\rightarrow\infty}\sum_{n=-N}^{N}\left\{ \intop_{2\pi n}^{2\pi\left(n+1\right)}d\phi\, e^{i(p-l+k)\phi}\right\} \\
 & = & \frac{1}{\sqrt{2\pi}}\intop_{-\frac{1}{2}}^{\frac{1}{2}}dl\, I_{0}(l)e^{\lambda_{0}(l)\tau}\sum_{p}c_{0}^{p}(l)\intop_{0}^{2\pi}d\phi\, e^{i(p-l+k)\phi}\lim_{N\rightarrow\infty}\sum_{n=-N}^{N}e^{2\pi i\left(p-l+k\right)n}\nonumber \\
 & = & \frac{1}{\sqrt{2\pi}}\intop_{-\frac{1}{2}}^{\frac{1}{2}}dl\, I_{0}(l)e^{\lambda_{0}(l)\tau}\sum_{p}c_{0}^{p}(l)\Xi\left(p-l+k\right)\intop_{0}^{2\pi}d\phi\, e^{i(p-l+k)\phi}.\nonumber 
\end{eqnarray}
As both $l$ and $k$ are from the first Brillouin zone and $p$ is
an integer the use of the Dirac comb function $\Xi(x)=\sum_{n\in Z}\delta(x-n\begin{scriptsize}
\begin{tiny}
•\begin{scriptsize}
\begin{tiny}
•\begin{scriptsize}
\begin{tiny}
•\begin{scriptsize}
\begin{tiny}
•
\end{tiny}
\end{scriptsize}
\end{tiny}
\end{scriptsize}
\end{tiny}
\end{scriptsize}
\end{tiny}
\end{scriptsize})$
implies $p=0$ and $k=l$ which leads to
\begin{equation}
F(k;\tau\to\infty)=\lambda_{0}(k)\tau+\ln\sqrt{2\pi}I_{0}(k)c_{0}^{0}(k).
\end{equation}
The second term depends on the initial state of the system and is
irrelevant in the long time limit. Zero-frequency $n$-th cumulant
of $v$ follows from \cite{Golubev10S}
\begin{equation}
C_{n}=\lim_{\tau\rightarrow\infty}\frac{(-i)^{n}}{\tau}\left.\frac{\partial^{n}F(k,\tau)}{\partial k^{n}}\right\rfloor _{k=0}=(-i)^{n}\left.\frac{\partial^{n}\lambda_{0}(k)}{\partial k^{n}}\right\rfloor _{k=0}.\label{eq:Cumulants}
\end{equation}

\bibliographystyle{apsrev4-1}


\end{document}